\documentclass{jfm}
\usepackage{graphicx}
\usepackage{epstopdf, epsfig}
\usepackage{amsmath, amsfonts}
\usepackage[x11names]{xcolor} 
\usepackage{tikz}
\usepackage{hyperref}

\usetikzlibrary{angles,quotes} 

\tikzset{
    cross/.pic = {
    \draw[rotate = 45] (-#1,0) -- (#1,0);
    \draw[rotate = 45] (0,-#1) -- (0, #1);
    }
} 
\tikzset{
  pics/carc/.style args={#1:#2:#3}{
    code={
      \draw[pic actions] (#1:#3) arc(#1:#2:#3);
    }
  }
} 

\usetikzlibrary{calc}
\DeclareRobustCommand\full     {\tikz[baseline=-0.6ex]\draw[thick] (0,0)--(0.5,0);}
\DeclareRobustCommand\fullfat  {\tikz[baseline=-0.6ex]\draw[line width=2pt] (0,0)--(0.5,0);}

\newcommand{\vc}[1]{\boldsymbol{\mathbf{#1}}} %
\newcommand{\df}{ \text{d} } 

\usepackage{nicematrix} 

\usepgfmodule{nonlineartransformations}
\usetikzlibrary{fpu} 
\newcommand{\PgfmathsetmacroFPU}[2]{\begingroup%
\pgfkeys{/pgf/fpu,/pgf/fpu/output format=fixed}%
\pgfmathsetmacro{#1}{#2}%
\pgfmathsmuggle#1\endgroup}
\tikzset{declare function={xtransformed(\x,\y)=\pgfkeysvalueof{/tikz/trafos/a}*
    sign(\x)*\y*\y;},
trafos/.cd,a/.initial=pi}
\makeatletter
\def\mytransformation{%
\PgfmathsetmacroFPU{\myy}{\pgf@y}%
\PgfmathsetmacroFPU{\myx}{\pgf@x+xtransformed(\pgf@x/1cm,\pgf@y/1cm)}
\pgf@x=\myx pt%
\pgf@y=\myy pt%
}
\makeatother

\shorttitle{The oscillating and vibrating cylinder}
\shortauthor{\'E. Spieser, S. Zhong and X. Zhang}

\title{The oscillating and vibrating cylinder: \\ a benchmark to study low Mach number aeroacoustics}

\author{\'Etienne Spieser\aff{1}\corresp{\email{meetiennes@ust.hk}} ,
  Siyang Zhong\aff{2}
 \and Xin Zhang\aff{1}}

\affiliation{\aff{1}Department of Mechanical and Aerospace Engineering, The Hong Kong University of Science and Technology, Clear Water Bay, Kowloon, Hong Kong SAR, China
\aff{2}Department of Aeronautical and Aviation Engineering, The Hong Kong Polytechnic University, Hung Hom, Kowloon, Hong Kong SAR, China}

\begin{document}

\maketitle

\begin{abstract}
Hydrodynamic and acoustic scales separate as the Mach number decreases, making the modelling of aeroacoustic phenomena singular in this flow regime.
The benchmark of the flow developing around an oscillating and vibrating cylinder is one of the scarce configuration that is fully analytically tractable, and is thus precious in the validation of new theories or solvers.
This work carefully derives the complete incompressible flow solution for this case, extending the axisymmetric results of the literature to more complex cylinder motion.
High-order finite element method solution of the incompressible Navier-Stokes equations provide a reference to validate the analytical formulae derived here.
Both analytical and numerical investigations agree on an independence of the configuration to the Reynolds number. Up to the critical Reynolds number, found to lie above $10^4$, the hydrodynamic solution of this configuration is solely governed by the Stokes number.
The exact expression of the sound radiated in the far field is computed by accounting for the cylinder scattering by means of tailored Green's functions.

\end{abstract}

\section{Introduction}

Acoustic analogies consist in a subtle rearrangement of the equations of fluid mechanics and aim to estimate the acoustic levels produced by a flow.
The prerequisite is that the problem of sound generation is decoupled from the problem of its propagation.
The study of the noise emitted by high-speed jets sparked this theory \citep{Lighthill_prsl52}, and up to now has been one of its focus leading to continuous research and improvements in acoustic analogies \citep{Phillips_jfm60, LilleyPlumbleeStrahleRuoDoak_1972, Goldstein_jfm03, SpieserBogeyBailly_jsv23}.
When the Mach number decreases, the speed at which hydrodynamic perturbations are convected becomes increasingly small with respect to the speed of sound. The scales associated with acoustic and hydrodynamic phenomena consequently depart from each other, and the physic of sound generation eventually becomes multiscale \citep{KlainermanMajda_cpam82, Alazard_arma06}.
This aeroacoustic singularity arising at low Mach numbers makes classical approaches used for compressible fluid dynamics not operable or inefficient \citep{MunzDumbserRoller_jcp07, HaackJinLiu_ccp12}.
Yet numerous important aeroacoustic applications are found at this flow regime such as noise in ventilation systems, underwater stealthiness, or community noise caused by drones. This is why, since early times, in parallel to these efforts to study high Mach number flows, low Mach number aeroacoustics has also been a subject of active research \citep{Ribner_utia62, Powell_jasa64, Crow_sam70, Mohring_jfm78}. 

Acoustic analogies neglect the feedback of sound on the vortex dynamics. They additionally assume a spatial separation between sound generation and sound propagation regions. What is more, they require the state variables to be split, in a more or less arbitrary way, between base flow and acoustic perturbations \citep{Fedorchenko_jfm97, Goldstein_jfm03, Goldstein_jfm05}.
These underlying modelling choices, inherent to the framework of acoustic analogies, led to a plurality of formulations, but also to several controversies \citep{McIntyre_jfm81, Fedorchenko_jsv00, Tam_jca02}.
Having a set of well-chosen reference cases is therefore essential to refine the different theoretical models, to select the most appropriate formulation and finally to gain insight in the sound generation mechanisms.
And indeed, the growth of the community has seen the development of a large number of benchmarks. 
This has been made possible thanks to sophisticated mathematical techniques \citep{CrightonDowlingFfowcsWilliamsHecklLeppington_92} along with the use of increasingly advanced numerical resolution methods and computing power \citep{HardinHussaini_93, LeleNichols_ptrsl14}.
The computational aeroacoustic workshops hosted by NASA (Tam \textit{et. al} \citeyear{HardinRistorcelliTam_NASA95, TamHardin_NASA97, HardinHuffTam_NASA00, DahlEnviaHuffTam_NASA04}) represented a significant contribution in this regard by allowing different theories to compete and encouraged scientific discussions.
The test cases described in these proceedings dealt with applications as various as the radiation of a baffled cylinder, follow-ups of \citep{TamWebb_jcp93} concerning the propagation of Gaussian pulses, the scattering of sound by multiple cylinders, or noise radiated from gusts impinging airfoils capitalising on the developments of \citet{Amiet_jsv86}.
The catalogue of aeroacoustic benchmarks is broader, and includes, but is not limited to, the acoustic radiation exhausting an unflanged duct with flow \citep{Munt_jfm77, ZhangChenMorfey_ija05}, the scattering of sound by a vortex \citep{ColoniusLeleMoin_jfm94}, the acoustic of a pair of contra-rotating vortices \citep{LeeKoo_aiaaj95, MitchellLeleMoin_jfm95}, the noise of a bidimensional mixing layer \cite{BogeyBaillyJuve_aiaaj00}, the sound radiated by a rectangular cavity \citep{Zhang_aiaaj95, GloerfeltBaillyJuve_jsv03}, the noise of a cylinder in flow \citep{CoxBrentnerRumsey_tcfd98, GloerfeltPerotBaillyJuve_jsv05, ZhengZhouZhongZhang_pof23} of a pair of cylinders \citep{BresFreedWesselsNoeltingPerot_pf12}, or, of a rod and an airfoil section in flow \citep{CasalinoJacobRoger_aiaaj03, JacobBoudetCasalinoMichard_tcfd05, GiretSengissenMoreauSanjoseJouhaud_aiaaj15}.

As already mentioned, the acoustic emission of low Mach number flows requires particular attention, as it involves multiscale physics for which the use of conventional techniques may be compromised.
To study this low Mach number limit, an aeroacoustic benchmark for which the hydrodynamic flow field as well as the acoustic field are fully analytically tractable seems ideal. 
It appears that the sound produced by an oscillating cylinder is one of the very few configurations which can be fully solved analytically.
This configuration was introduced by \citet{Meecham_jasa65} to study the sound sources arising in boundary layers. In this work, the surface and volume contribution to the radiated noise are estimated by means of dimendional analysis and found to be of comparable order, the diffraction caused by the cylinder is omitted. What is more no details for the derivation of this incompressible flow solution were provided. 
This topic was revisited by \citet{Lauvstad_jasa68}, who carefully derived the compressible solution for this benchmark using a matched asymptotic expansion in the Low Mach number limit. Lauvstad took into account the cylinder scattering effect and proposed a solution for the sound radiated in the far field.
Shortly after, in a joint contribution, \citet{LauvstadMeecham_jsv69} recast Lighthill's and Curle's theories in covariant form using tensor calculus. The application of their formulation to the case of the oscillating cylinder indicated that the surface and volume sound sources cancel each other out perfectly, leading to zero net sound production.
Later on, contradicting this result, \citet{Meecham_jsp73} found a non-vanishing acoustic radiation for this benchmark using his simple-source theory.
This benchmark has been revisited by \citet{MorfeySorokinGabard_jfm12} to illustrate the effect of non-rigid surfaces in viscosity-induced sound generation processes. 
These authors introduced scalar and vector potentials to solve the compressible flow problem. 
Compared to previous formulations, these derivations are more general, do not assume axisymmetry of the problem, allow normal velocities of the cylinder and do not assume the Mach number to be infinitesimally small. Yet, because the non-linear terms were dropped in \citet{MorfeySorokinGabard_jfm12}, the oscillating cylinder is found to radiate no sound when described with this framework.
This was pointed out later by \citet{MorfeySorokinWright_jsv22}. 
In their more recent study, \citet{MorfeySorokinWright_jsv22} identified the flow shear as well as the viscous dissipation as the two mechanisms responsible for sound generation. They both correspond to non-linear volume sound sources.
These authors derive the velocity field from the vorticity equation and account for the scattering of the cylinder with tailored Green's functions. Their aeroacoustic analysis however is restricted to large Stokes numbers, and unlike \citet{MorfeySorokinGabard_jfm12} cannot account for higher-order modes of oscillation or vibration of the cylinder. 

The flow around an oscillating cylinder is one of the scarce fully analytically tractable configuration that enables the study of low Mach number aeroacoustics, yet contradictory results are found in the literature.
The present study intends to thoroughly review this benchmark and to extend its range of validity.
This paper is organised as follows: An example of the usage of Lighthill's acoustic analogy informed with a naive incompressible sound source is given in \S \ref{txt:failureOfLighthill} to introduce some subtleties of low Mach number aeroacoustics and advocate for the need of benchmark case at this flow regime.
In \S \ref{txt:generalSolutionCylinder} the general equations governing the incompressible flow dynamic around an oscillating and vibrating cylinder are carefully derived with a strategy which is valid for more complex surface motion than in \citet{MorfeySorokinWright_jsv22}.
The solution is then particularised in \S \ref{txt:undeformableCylinder} for the case of a purely oscillating cylinder. The solution's validity over a wide range of Stoke's number is then verified numerically with a finite element resolution of the problem.
Then in \S \ref{txt:acousticFarFieldSol}, the acoustic far-field solution is computed with help of the low Mach number acoustic analogies of Powell and of Ribner.
This work does not investigated the noise stemming from viscous dissipation.
Green's functions tailored to the geometry problem are considered to account for the reflection and diffraction of sound wave on the cylinder surface.
A discussion and set of final remarks conclude this study.

\section{Lighthill's acoustic analogy in the low Mach number limit}
\label{txt:failureOfLighthill}

In principle, different base flow can be used to describe sources of sound \citep[{\S} 3.3]{Goldstein_jfm03, WangFreundLele_ar06}, but care must be taken to accordingly select a consistent acoustic analogy. This is highlighted here, considering a naive implementation of Lighthill's acoustic analogy with incompressible flow data.
Let the density fluctuation associated with acoustics be introduce as $\rho'$, and let the variables $\vc{u}_i$, $p_i$ and $\rho_i$, indexed by ``$_i$'', correspond to the incompressible base flow velocity, pressure and density respectively, so that $\nabla \cdot \vc{u}_i = 0$ and $\rho_i = Cst$. Lighthill's acoustic analogy then reads
\begin{equation}
    \dfrac{\partial^2 \rho'}{\partial t^2} - a_0^2 \Delta \rho' = \nabla \cdot \nabla \cdot  \vc{T} \ ,
    \label{eq:Lighthill's_analogy}
\end{equation}
with
\begin{equation}
    \vc{T} = \underbrace{\rho_i \vc{u}_i \otimes \vc{u}_i}_{(a)} + \underbrace{(p_i - a_0^2 \rho_i) \vc{I}_d}_{(b)} - \underbrace{\rho_i \nu \left( (\nabla \vc{u}_i) + (\nabla \vc{u}_i)^T\right)}_{(c)}
    \ ,
    \label{eq:Lighthill's_source}
\end{equation}
where the stress tensor $\vc{T}$ is informed with the incompressible flow data, and let its terms be grouped and labelled as in the above.
The speed of sound $a_0$ is arbitrarily introduced and taken constant as in \citet{Lighthill_prsl52}, $\nu$ corresponds to the kinematic viscosity of the fluid.

First, using vector calculus, it is straightforward to show that the contribution of the term $(c)$ vanishes when the velocity field is solenoidal\footnote{Recalling that for all $\vc{v}$, $\nabla \cdot (\nabla \vc{v})= \Delta \vc{v}$ with $\nabla \cdot \left(\Delta \vc{v} \right) = \Delta \left( \nabla \cdot \vc{v} \right) $ and $\nabla \cdot \left( (\nabla \vc{v})^T\right) = \nabla (\nabla \cdot \vc{v})$.}.
An equation of state $\delta p = \left( \partial p/\partial \rho\right)_{s} \delta \rho + \left( \partial p/\partial s\right)_{\rho} \delta s$ is sometimes used to justify that the term $(b)$ corresponds to a sound source associated with entropy and is negligible in most flow noise application under the assumption of homentropicity. 
However this state equation ceases to be valid in the incompressible limit, for which $\delta \rho \rightarrow 0$ and $\left( \partial p/\partial \rho\right)_{s} \rightarrow \infty$. Another relation must be used to interpret the term $(b)$ in the low Mach number limit. This is achieved with the momentum equation,
\begin{equation}
    \dfrac{\partial \vc{u}_i}{\partial t} + \nabla \cdot (\vc{u}_i \otimes \vc{u}_i) + \dfrac{\nabla p_i}{\rho_i} - \nu \Delta \vc{u}_i = \vc{0}
    \ .
\end{equation}
It is straightforward to see that the divergence of this equation leads to well-known Poisson's equation for the incompressible pressure $p_i$,
\begin{equation}
    \Delta p_i = - \nabla \cdot \nabla \cdot (\rho_i \vc{u}_i \otimes \vc{u}_i)
    \ .
    \label{eq:Poisson's_eq}
\end{equation}
Since in the present framework $a_0^2 \rho_i$ is constant, the double divergence of terms $(a)$ and $(b)$ of equation (\ref{eq:Lighthill's_source}) cancel each other out. The sound source of Lighthill's acoustic analogy $\nabla \cdot \nabla \cdot \vc{T}$ is hence identically null when an incompressible base flow is used to model the sound source.

This simple example recalls that standard aeroacoustic methods may be compromised in the low Mach number limit and deserve a special attention.
This issue is well identified in the literature \citep{Lighthill_prsl52, Crow_sam70}, and when the Mach number tends toward zero, Lighthill's stress tensor \textit{must} (and not \textit{can}) be simplified into,
\begin{equation}
    \vc{T} \approx \rho_i \vc{u}_i \otimes \vc{u}_i    
    \ .
    \label{eq:Lighthill_incompSource}
\end{equation}
Section {\S} \ref{txt:Powell_Ac_Analogy} illustrates that this is consistent with other low Mach number acoustic analogies of the literature \citep{Ribner_utia62, Powell_jasa64}.
This subtlety and others, such as the erroneous prediction of pressure disturbances obtained from incompressible calculations when solids are not acoustically compact, (e.g. see \citet[{\S} 5.6]{GleggDevenport_17}), encourage the use of analytically tractable benchmark to deepen the understanding of noise generation mechanisms and design more suitable models.

\section{Solution of the flow around an oscillating and vibrating cylinder}
\label{txt:generalSolutionCylinder}

\begin{figure}
    \centering
    \def\r0{2} 
    \def\thet0{66} 
    \def\d0{3} 
    \def\ang0{-25} 
    \begin{tikzpicture}
        \begin{scope}
            \pgftransformnonlinear{\mytransformation}
            \fill[fill=purple!20!cyan!15!white!100] (0,0) circle[radius=\r0];
            \draw  (0,0) circle[radius=\r0] node [yshift=-13mm] {Solid};
            \draw  (3.5,0) node [yshift=+8mm] {Fluid};
        \end{scope}
        \draw   ({\d0*cos(\ang0)}, {\d0*sin(\ang0)-0.2}) node {$\vc{x}$};
        \path ({\d0*cos(\ang0)}, {\d0*sin(\ang0)}) pic[rotate = 45] {cross=1mm};
        \draw[dashed]   (0,0) -- (\r0,0);
        \draw[-latex]   (0,0) -- node [above] {$r$} ({\d0*cos(\ang0)}, {\d0*sin(\ang0)}); 
        \draw[-latex]   ({\d0*cos(\ang0)}, {\d0*sin(\ang0)}) -- node [right, xshift=0.2cm] {\small$\vc{e}_r$} ({\d0*cos(\ang0)+0.5*cos(\ang0)}, {\d0*sin(\ang0) + 0.5*sin(\ang0)}); 
        \draw[-latex]   ({\d0*cos(\ang0)}, {\d0*sin(\ang0)}) -- node [left, yshift=0.2cm] {\small$\vc{e}_\theta$} ({\d0*cos(\ang0)-0.5*sin(\ang0)}, {\d0*sin(\ang0)+0.5*cos(\ang0)}); 
        \draw[dashed]   (0,0) -- node [left] {$r_0(\theta,t)$} ({cos(\thet0)*\r0*1.07}, {sin(\thet0)*\r0*1.07}) ;
        \draw[-latex]   ({cos(\thet0)*\r0*1.07}, {sin(\thet0)*\r0*1.07}) --  ({cos(\thet0)*\r0 - 1.2*sin(\thet0)}, {sin(\thet0)*\r0 + 1.2*cos(\thet0)}); 
        \draw ({cos(\thet0)*\r0-1.1}, {sin(\thet0)*\r0+0.65}) node {$u_\theta(r_0,\theta,t)$};
        \draw[-latex]   ({cos(\thet0)*\r0*1.07}, {sin(\thet0)*\r0*1.07}) --  ({cos(\thet0)*\r0 + 0.5*cos(\thet0)}, {sin(\thet0)*\r0 + 0.5*sin(\thet0)}); 
        \draw ({cos(\thet0)*\r0+1.0}, {sin(\thet0)*\r0+0.3}) node {$u_r(r_0,\theta,t)$};
        \path (0,0) pic[-latex]{carc=0:\thet0:0.5cm}; 
        \draw (0.45,0.5) node {$\theta$};
    \end{tikzpicture}
    \caption{Sketch of a deformed cylinder undergoing periodic oscillations and vibrations.}
    \label{fig:sketchGeneralCase}
\end{figure}
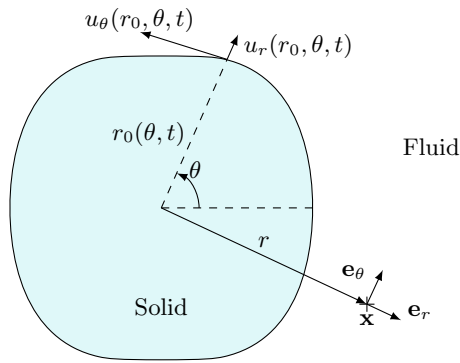

The problem to be solved is to determine the flow around an oscillating and vibrating cylinder whose wall is subjected to forced periodic deformations and rotations. The solution of this problem is sought assuming the flow is incompressible.
The radiation of sound is to be computed in a subsequent step.
A polar coordinate system ($\vc{e}_r$,$\vc{e}_\theta$) defined by a radius $r$ and a polar angle $\theta$ is used, so that the radial and azimuthal velocity fields are labelled as $u_r$ and $u_\theta$. 
Due to the forced oscillating and vibrating motion, the cylinder radius $r_0$ may vary with time and with the polar angle as sketched in figure \ref{fig:sketchGeneralCase}.
The deformations and rotations applied to the cylinder's surface are supposed small enough so that non-linearities do not play a significant role in the determination of the velocity field. This assumption will be verified numerically in {\S} \ref{txt:numericalMethod} on some specific examples.


\subsection{General solution to the vorticity equation}
\label{eq:GeneralSolVorticityEq}

The vorticity field $\omega$ for a two-dimensional problem is defined as a scalar field pointing out of the plane. If $\vc{u} = u_r \vc{e}_r + u_\theta \vc{e}_\theta$ is the general expression of the fluid velocity, let $\vc{e}_z$ be the out of plane direction, such that,
\begin{equation}
    \omega = (\nabla \times \vc{u}) \cdot \vc{e}_z = \dfrac{1}{r} \left( \dfrac{\partial (r u_\theta)}{\partial r} - \dfrac{\partial u_r}{\partial \theta} \right)
    \ .
\end{equation}
The incompressible flow problem is written in terms of the vorticity field, as in \citep{MorfeySorokinWright_jsv22}. This strategy is chosen because fairly complex bidimensional flow features around the cylinder can fully be described by the following simple scalar second order equation,
\begin{equation}
    \dfrac{\partial \omega}{\partial t} - \nu \Delta \omega = 0
    \ .\label{eq:homogeneous_vorticity_eq_NABLA}
\end{equation}
Here again, $\nu$ refers to the kinematic viscosity of the fluid. This approach did not require any particular assumption apart from the linearity of the velocity field and incompressibility.
To account for higher order modes of the cylinder deformation (discarded in all previous studies except for \citet{MorfeySorokinGabard_jfm12}), the azimuthal dependency in the partial differential equation for the vorticity is kept,
\begin{equation}
    \dfrac{\partial \omega}{\partial t} - \nu \left[ \dfrac{1}{r} \dfrac{\partial}{\partial r}\left( r \dfrac{\partial \omega }{\partial r}\right) + \dfrac{1}{r^2}\dfrac{\partial^2 \omega}{\partial \theta^2} \right] = 0 
    \ .
    \label{eq:homogeneous_vorticity_eq}
\end{equation}
The solution is sought by separation of variables, postulating $\omega(r,\theta,t) = f(r)g(\theta)h(t)$. Equation (\ref{eq:homogeneous_vorticity_eq}) then becomes,
\begin{equation}
    \dfrac{1}{\nu}\dfrac{h'}{h} = \dfrac{rf''+f'}{rf} + \dfrac{1}{r^2} \dfrac{g''}{g}
    \ .
\end{equation}
The left-hand side depends only on time, while the right-hand side is time independent.  Let $-\lambda$ be this time-independent constant, it follows that the previous equation can be recast into,
\begin{equation}
    \dfrac{r^2f''+rf'}{f} + \lambda r^2 = - \dfrac{g''}{g}
    \ .
\end{equation}
Similarly, the left- and right-hand side again depend on different variables and therefore are constant. Let $n^2$ be this constant, equation (\ref{eq:homogeneous_vorticity_eq}) can then be transformed into a set of three ordinary differential equations whose solution are known,
\begin{equation}
    \left\{
    \begin{array}{ll}
        h' + \nu \lambda h = 0 & \rightarrow h(t) \propto e^{-\nu \lambda t} \\
        g'' + n^2g = 0 & \rightarrow g(\theta) \propto \alpha_n e^{in\theta} + \beta_n e^{-in\theta} \\
        r^2f''+rf'+ (\lambda r^2-n^2)f = 0 & \rightarrow f(r) \propto \zeta_n J_n(\sqrt{\lambda}r) + \eta_n Y_n(\sqrt{\lambda}r)
    \end{array}
    \right.
\end{equation}

\subsubsection{Finite solution at large time}

For the solution not to diverge with increasing time, Re$(\lambda)\ge0$.

\subsubsection{Angular periodicity}

The geometry of the problem imposes an angular periodicity of the solution, hence $g(\theta + 2\pi) = g(\theta)$. From this condition, $n\in\mathbb{Z}$.

\subsubsection{Finite solution at large radius}

When $\lambda \notin \mathbb{R}$, then $J_n$ and $Y_n$ may diverge for large arguments. Asymptotic expansion for Bessel functions\cite[{\S}8.3.1]{StoneGoldbart_09} are used to determine the condition on $\zeta_n$ and $\eta_n$ for the solution to remain finite at large radii.
\begin{equation}
    \begin{array}{c}
        J_n(z) \sim \sqrt{\dfrac{2}{\pi z}} \cos\left(z-\dfrac{(2n+1)\pi}{4}\right) \left( 1+O\left( \dfrac{1}{z} \right) \right) 
        \ ,\\[3ex]
        Y_n(z) \sim \sqrt{\dfrac{2}{\pi z}} \sin\left(z-\dfrac{(2n+1)\pi}{4}\right) \left( 1+O\left( \dfrac{1}{z} \right) \right) 
        \ .
    \end{array}    
\end{equation}
For $\zeta_n J_n(\sqrt{\lambda}r) + \eta_n Y_n(\sqrt{\lambda}r)$ to remain finite when $r\rightarrow \infty$, $\eta_n/\zeta_n = i$ must be chosen when Im$(\lambda) \ge 0$ and $\eta_n/\zeta_n = -i$ when Im$(\lambda) \le 0$.
Hence, the $\zeta_n J_n(\sqrt{\lambda}r) + \eta_n Y_n(\sqrt{\lambda}r)$ sum corresponds to the Hankel function of the first or of the second kind such that,
\begin{equation}
    f(r) \propto \left\{
    \begin{array}{ll}
         H_n^{(1)}(\sqrt{\lambda}r) & \text{when Im$(\lambda) \ge 0$}
         \ ,\\ 
         H_n^{(2)}(\sqrt{\lambda}r) & \text{when Im$(\lambda) \le 0$} 
         \ .\\ 
    \end{array}
    \right.
\end{equation}
As the case study {\S} \ref{txt:undeformableCylinder}  will show, $\lambda$ takes values such that the Hankel function of the first kind $H_n^{(1)}$ must be chosen. If a different configuration were to arise in practice, it would be easy to modify this expression in favor of the Hankel function of the second kind $H_n^{(2)}$.

\subsubsection{Final expression}

The general solution for diffusion equation of the vorticity, equation (\ref{eq:homogeneous_vorticity_eq}), is hence a linear combination of $\left( \alpha_n e^{in\theta} + \beta_n e^{-in\theta}  \right) H_n^{(1)}(\sqrt{\lambda}r) e^{-\nu \lambda t}$ terms, for continuous values of $\lambda$ verifying Re$(\lambda)\ge0$ and Im$(\lambda) \ge 0$, and discrete values of $n\in\mathbb{Z}$. From the reflection formula of Bessel and Hankel functions $H_{-n}^{(1)}(x) = (-1)^nH_n^{(1)}(x)$, the sum over $\mathbb{Z}$ can thus be recast into the following sum over $\mathbb{N}$, leading to,
\begin{equation}
    \omega(r,\theta,t) = \sum_{n \in \mathbb{N}} 
    \left( \alpha_{n} e^{in\theta}  + \beta_{n} e^{-in\theta} \right) 
    H_n^{(1)}(\sqrt{\lambda}r)
    e^{- \nu \lambda t}
    \ .
    \label{eq:generalSol_w}
\end{equation}

\subsection{Taking into account the time-varying boundary conditions}
\label{txt:takingAcountBC}

The boundary conditions are most easily enforced by considering the streamfunction $\psi$ from which the velocity field derives. In polar coordinates, $\psi(r,\theta,t)$ is defined so that,
$\vc{u} = \left( u_r, u_\theta \right)^T = \left( \frac{1}{r} \frac{\partial \psi}{\partial \theta}, -\frac{\partial \psi}{\partial r} \right)^T$
 where $\psi$ is deduced from the vorticity field $\omega$ from,
\begin{equation}
    \Delta \psi (r,\theta,t) = - \omega(r,\theta,t) H(r-r_0) \hspace{0.5cm} \text{and} \hspace{0.5cm} 
    \left\{
    \begin{array}{l}
        \left. \dfrac{\partial \psi}{\partial \theta} \right|_{r=r_0}= r_0\, u_r(r_0, \theta, t ) 
        \ ,\\
        \left. \dfrac{\partial \psi}{\partial r} \right|_{r=r_0}=-u_\theta(r_0, \theta ,t) 
        \ ,\\
        \displaystyle \lim_{r\rightarrow\infty} \left(\psi\right) = 0
        \ .
    \end{array}
    \right.
    \label{eq:inhomogeneous_PDE_on_psi}
\end{equation}
In the above, the Heaviside function $H(r-r_0)$ is used to select the contribution of the vorticity around the cylinder.
Equation (\ref{eq:inhomogeneous_PDE_on_psi}) is an inhomogeneous partial differential equation with time-varying boundary conditions. 
The solution can be computed by splitting the vector potential $\psi$ into a free-field contribution $\psi_G$ taking account of the vorticity distribution by means of Green's representation theorem, and of a homogeneous solution that satisfies the cylinder boundary condition $\psi_H$. That is, equation (\ref{eq:inhomogeneous_PDE_on_psi}) may be recast into,
\begin{equation}
    \psi = \psi_G + \psi_H
    \ ,
    \label{eq:splitOfPsi}
\end{equation}
where 
\begin{equation}
    \Delta \psi_G(r,\theta,t) = -\omega(r,\theta,t)\, H(r-r_0)
    \hspace{0.5cm} \text{and} \hspace{0.5cm} 
    \Delta \psi_H (r,\theta,t) = 0 
    \label{eq:Poisson_Laplace_eqs_for_psi}
\end{equation}
with the boundary conditions on the cylinder radius,
\begin{equation}
    \left. \dfrac{\partial \psi_H}{\partial r} \right|_{r=r_0}=- \left[ u_\theta + \dfrac{\partial \psi_G}{\partial r} \right]_{r=r_0}
    \hspace{0.5cm} \text{and} \hspace{0.5cm}
    \left. \dfrac{\partial \psi_H}{\partial \theta} \right|_{r=r_0}= \left[ r\, u_r - \dfrac{\partial \psi_G}{\partial \theta} \right]_{r=r_0}
\end{equation}
and in the far field,
\begin{equation}
    \displaystyle \lim_{r\rightarrow\infty} \left(\psi_H\right) = - \lim_{r\rightarrow\infty} \left(\psi_G\right)
    \ .
\end{equation}

\subsubsection{Solution to the homogeneous problem $\psi_H$}

As for the vorticity equation, the solution to this homogeneous equation $\psi_H(r,\theta)$ is sought by introducing $\psi_H(r,\theta) = p(r)q(\theta)$ into the Laplace equation $\Delta \psi_H = 0$. One then obtains,
\begin{equation}
    \dfrac{r^2p''+rp'}{p} = -\dfrac{q''}{q}
    \ .
\end{equation}
The right- and left-hand side of this equation depend on distinct variables, thus the members of this equation are constant. Let $m^2$ be this constant. It follows,
\begin{equation}
    \left\{
    \begin{array}{ll}
        q'' + m^2q = 0 & \rightarrow  q(\theta) \propto \sigma_m e^{im\theta} + \tau_m e^{-im\theta}
        \ ,\\[2ex]
        r^2p''+rp'-m^2p = 0 & \rightarrow p(r) \propto 
        \left\{
        \begin{array}{ll}
             \ln(r) & \text{when $m = 0$} 
             \ ,\\
             r^{\pm m} & \text{when $m \neq 0$}
             \ .
        \end{array} \right.
    \end{array}
    \right.
\end{equation}
From the angular periodicity of the problem, again $m\in\mathbb{Z}$. Whereas, from the condition of finite solution at large radius, for $m\neq0$, $p(r) \propto r^{-|m|}$. Thus, the sum over $m$ can again be used to form the general solution,
\begin{equation}
    \psi_H(r,\theta,t) = (\sigma_0 + \tau_0) \ln\left( \dfrac{r}{r_0} \right)  + \sum_{m\in\mathbb{N^*}} \left(\sigma_m e^{im\theta} + \tau_m e^{-im\theta} \right) \left(\dfrac{r_0}{r}\right)^{m}   
    \ ,
    \label{eq:homogeneous_solution_psi_H}
\end{equation}
where the radius of the cylinder $r_0$ has been added to non-dimensionalise the solution, and where the coefficients $\sigma_m$ and $\tau_m$ are constants of $r$ and $\theta$, but may vary with time.

\subsubsection{Solution to the free-field problem $\psi_G$}

Let us find $\psi_G$ with help of Green's function $G_{r_s,\theta_s}$ associated with Poisson's equation,
\begin{equation}
    \Delta G_{r_s,\theta_s}(r,\theta) = \dfrac{1}{r} \delta (r-r_s) \delta (\theta-\theta_s)
    \ .
    \label{eq:Poisson's_Green's_eq}
\end{equation}
Its solution in polar coordinate system expressed over $m\in\mathbb{Z}$ \cite[{\S} 3.10]{Jackson_99} \cite[Example 5.1.1]{Duffy_15}
, is recast into a sum over $\mathbb{N}$ such that,
\begin{equation}
    G_{r_s,\theta_s}(r,\theta) = \dfrac{1}{2\pi}\ln\left(\dfrac{r_>}{r_0}\right) - \sum_{m\in\mathbb{N}^*} \dfrac{1}{4\pi m}
    \left( e^{im(\theta-\theta_s)} + e^{-im(\theta-\theta_s)} \right)
    \left(\dfrac{r_<}{r_>} \right)^{m}
    \ ,
    \label{eq:Poisson's_Green's_fct}
\end{equation}
where $r_< = \min(r,r_s)$ and $r_> = \max(r,r_s)$. Note the resemblance of previous Green's function is the solution given in equation (\ref{eq:homogeneous_solution_psi_H}).
In this free-field formulation, the solution for the streamfunction $\psi_G$ vanishes in the far field and no contour integral have to be taken into account in Green's representation formula. Hence, $\psi_G$ is simply obtained from,
\begin{equation}
    \psi_G(r,\theta,t) = -\displaystyle \int_0^{2\pi} \int_{r_0}^\infty G_{r_s,\theta_s}(r,\theta)\, \omega(r_s,\theta_s,t)\, r_s\, \df r_s \, \df \theta_s
    \label{eq:beforeAssumptionOfSmallDeformations}
\end{equation}
where the effect of the Heaviside function in the source term of Poisson's equation given in (\ref{eq:Poisson_Laplace_eqs_for_psi}) has been accounted the integral range.
Replacing $G_{r_s,\theta_s}$ and $\omega$ by their respective expressions (\ref{eq:Poisson's_Green's_fct}) and (\ref{eq:generalSol_w}), and recalling that 
\begin{equation}
    \int_0^{2\pi} e^{i(m-n)\theta_s } \df \theta_s = 2\pi \delta_{m,n} 
    \ ,
\end{equation}
it follows from the angular integration\footnote{Here it has been assumed that the deformations of the surface are small with respect to the cylinder mean radius $r_0$. This enabled us in equation (\ref{eq:beforeAssumptionOfSmallDeformations}) to perform the integration over $\theta_s$ independently of the integration over $r_s$.
If the deformations were not small, equation (\ref{eq:beforeAssumptionOfSmallDeformations}) could not in general be further simplified, and the integral bound would have to be informed with $r_0(\theta_s, t)$, which is a known parameter of the problem.},
\begin{equation}
\begin{array}{rl}
    \psi_G(r,\theta,t) =&\hspace{-0.2cm} \displaystyle 
    -(\alpha_0 + \beta_0)e^{ - \nu \lambda t}
    \int_{r_0}^\infty r_s\ln\left(\dfrac{r_>}{r_0}\right) H_0^{(1)}(\sqrt{\lambda}r_s)
    \df r_s\\
    &\hspace{-0.2cm} \displaystyle + \sum_{n \in \mathbb{N}^*} 
    \left( \alpha_n e^{in\theta} + \beta_{n} e^{-in\theta} \right) e^{ - \nu \lambda t}
    \int_{r_0}^\infty
    \dfrac{r_s}{2n} \left(\dfrac{r_<}{r_>} \right)^{n} H_n^{(1)}(\sqrt{\lambda}r_s)
    \df r_s
    \ ,
\end{array}
\end{equation}
And after splitting the integration bounds,
\begin{equation}
\begin{array}{rl}
    \hspace{-1cm} \psi_G(r,\theta,t) =&\hspace{-0.2cm} \displaystyle 
    -(\alpha_0 + \beta_0)e^{ - \nu \lambda t}
    \left[
    \int_{r_0}^r r_s\ln\left(\dfrac{r}{r_0}\right) H_0^{(1)}(\sqrt{\lambda}r_s)
    \df r_s
    + \int_{r}^\infty r_s\ln\left(\dfrac{r_s}{r_0}\right) H_0^{(1)}(\sqrt{\lambda}r_s)
    \df r_s
    \right]\\
    &\hspace{-2.2cm} \displaystyle + \sum_{n \in \mathbb{N}^*} 
    \left( \alpha_n e^{in\theta} + \beta_{n} e^{-in\theta} \right) e^{ - \nu \lambda t}
    \left[
    \int_{r_0}^r
    \dfrac{r_s}{2n} \left(\dfrac{r_s}{r} \right)^{n} H_n^{(1)}(\sqrt{\lambda}r_s)
    \df r_s
    + \int_{r}^\infty
    \dfrac{r_s}{2n} \left(\dfrac{r}{r_s} \right)^{n} H_n^{(1)}(\sqrt{\lambda}r_s)
    \df r_s
    \right]
    \ .
\end{array}
\label{eq:freefield_solution_psi_G}
\end{equation}

\subsubsection{Matching the radial velocity on the cylinder}

This condition is
\begin{equation}
    \dfrac{\partial \psi_H}{\partial \theta} (r_0, \theta ,t) = r_0 u_r(r_0,\theta,t) - \dfrac{\partial \psi_G}{\partial \theta}(r_0, \theta ,t)
    \ ,
    \label{eq:generalEq--radialVel}
\end{equation}
with the expression of $\psi_H$ given in equation (\ref{eq:homogeneous_solution_psi_H}) leading to,
\begin{equation}
    \dfrac{\partial \psi_H}{\partial \theta} (r_0, \theta, t) = 
     \sum_{m\in\mathbb{N^*}} im \left(\sigma_m e^{im\theta} - \tau_m e^{-im\theta} \right) 
     \label{eq:generalEq--radialVel_psiH}
\end{equation}
and the expression of $\psi_G$ given in equation (\ref{eq:freefield_solution_psi_G}) furnishing,
\begin{equation}
    \dfrac{\partial \psi_G}{\partial \theta} (r_0, \theta, t) = 
    \sum_{n \in \mathbb{N}^*} 
    \dfrac{i r_0}{2}
    \left( \alpha_n e^{in\theta} - \beta_{n} e^{-in\theta} \right) e^{ - \nu \lambda t}
    \int_{r_0}^\infty
    \left(\dfrac{r_0}{r_s} \right)^{n-1} H_n^{(1)}(\sqrt{\lambda}r_s)
    \df r_s
    \ .
    \label{eq:generalEq--radialVel_psiG}
\end{equation}

\subsubsection{Matching the azimuthal velocity on the cylinder}

This other condition on the cylinder surface is
\begin{equation}
    \dfrac{\partial \psi_H}{\partial r} (r_0, \theta ,t) =-u_\theta(r_0, \theta ,t) - \dfrac{\partial \psi_G}{\partial r} (r_0, \theta ,t)
    \ ,
    \label{eq:generalEq--azimuthalVel}
\end{equation}
where, from the expression of $\psi_H$ given in equation (\ref{eq:homogeneous_solution_psi_H}), 
\begin{equation}
    \dfrac{\partial \psi_H}{\partial r}(r_0,\theta,t) = 
    \dfrac{\sigma_0 + \tau_0}{r_0}  - \sum_{m\in\mathbb{N^*}} \dfrac{m}{r_0} \left(\sigma_m e^{im\theta} + \tau_m e^{-im\theta} \right) 
    \label{eq:generalEq--azimuthalVel_psiH}
\end{equation}
The radial derivative of $\psi_G$ is computed from equation (\ref{eq:freefield_solution_psi_G}) using Leibniz's integral rule. One finally obtains
\begin{equation}
    \dfrac{\partial \psi_G}{\partial r}(r_0,\theta,t) = 
    \sum_{n \in \mathbb{N}^*} 
    \left( \alpha_n e^{in\theta} + \beta_{n} e^{-in\theta} \right) \dfrac{e^{ - \nu \lambda t}}{2}
    \int_{r_0}^\infty
    \left(\dfrac{r_0}{r_s} \right)^{n-1} H_n^{(1)}(\sqrt{\lambda}r_s)
    \df r_s
    \ .
    \label{eq:generalEq--azimuthalVel_psiG}
\end{equation}

\subsubsection{Matching the solution in the far field}

This far-field condition,
\begin{equation}
    \lim_{r\rightarrow\infty}\left(\psi_G + \psi_H\right) = 0
\end{equation}
is obtained by matching the leading order of $\psi_G$ and $\psi_H$ which are the only non-zero contribution of these fields at large radiuses. Finally,
\begin{equation}
    \dfrac{\sigma_0 + \tau_0}{r_0}
    = (\alpha_0 + \beta_0) e^{-\nu \lambda t} \int_{r_0}^\infty
    \left(\dfrac{r_0}{r_s}\right)^{-1} H_0^{(1)}(\sqrt{\lambda}r_s) 
    \df r_s
    \label{eq:generalEq--farField}
\end{equation}

The derivations presented here enable to model non-axisymmetric displacement of the cylinder surface, for which the skin is possibly stretchable (see appendix \ref{txt:CylinderFlexiSkin}) or pulsating as in \cite{MorfeySorokinGabard_jfm12}.
The next section illustrates how the equations for the radial velocity (\ref{eq:generalEq--radialVel}), (\ref{eq:generalEq--radialVel_psiH}), (\ref{eq:generalEq--radialVel_psiG}) and those for the azimuthal velocity (\ref{eq:generalEq--azimuthalVel}), (\ref{eq:generalEq--azimuthalVel_psiH}), (\ref{eq:generalEq--azimuthalVel_psiG}) together with the far-field condition (\ref{eq:generalEq--farField}) can be used and particularised to determine the constants appearing in equations (\ref{eq:homogeneous_solution_psi_H}) and (\ref{eq:freefield_solution_psi_G}) and thereby determine the velocity vector potential, given in equation (\ref{eq:splitOfPsi}). From this streamfunction, the velocity and the pressure fields are computed and aeroacoustic predictions are made.

\section{Application to the undeformable oscillating cylinder}
\label{txt:undeformableCylinder}

\begin{figure}
    \centering
    \def\r0{2} 
    \def\thet0{66} 
    \def\d0{3} 
    \def\ang0{-25} 
    \begin{tikzpicture}
        \fill[fill=purple!20!cyan!15!white!100] (0,0) circle[radius=\r0];
        \draw  (0,0) circle[radius=\r0] node [yshift=-13mm] {Solid};
        \draw  (3.5,0) node [yshift=+8mm] {Fluid};
        \draw   ({\d0*cos(\ang0)}, {\d0*sin(\ang0)-0.2}) node {$\vc{x}$};
        \path ({\d0*cos(\ang0)}, {\d0*sin(\ang0)}) pic[rotate = 45] {cross=1mm};
        \draw[dashed]   (0,0) -- (\r0,0);
        \draw[-latex]   (0,0) -- node [above] {$r$} ({\d0*cos(\ang0)}, {\d0*sin(\ang0)}); 
        \draw[-latex]   ({\d0*cos(\ang0)}, {\d0*sin(\ang0)}) -- node [right, xshift=0.2cm] {\small$\vc{e}_r$} ({\d0*cos(\ang0)+0.5*cos(\ang0)}, {\d0*sin(\ang0) + 0.5*sin(\ang0)}); 
        \draw[-latex]   ({\d0*cos(\ang0)}, {\d0*sin(\ang0)}) -- node [left, yshift=0.2cm] {\small$\vc{e}_\theta$} ({\d0*cos(\ang0)-0.5*sin(\ang0)}, {\d0*sin(\ang0)+0.5*cos(\ang0)}); 
        \draw[dashed]   (0,0) -- node [left] {$r_0$} ({cos(\thet0)*\r0}, {sin(\thet0)*\r0}) ;
        \draw[-latex]   ({cos(\thet0)*\r0}, {sin(\thet0)*\r0}) --  ({cos(\thet0)*\r0 - 1.2*sin(\thet0)}, {sin(\thet0)*\r0 + 1.2*cos(\thet0)}); 
        \draw ({cos(\thet0)*\r0}, {sin(\thet0)*\r0+0.5}) node {$u_\theta(r_0,\theta, t)$};
        \path (0,0) pic[-latex]{carc=0:\thet0:0.5cm}; 
        \draw (0.45,0.5) node {$\theta$};
    \end{tikzpicture}
    \caption{Sketch of a cylinder in purely oscillatory motion.}
    \label{fig:sketchOscillatingCase}
\end{figure}
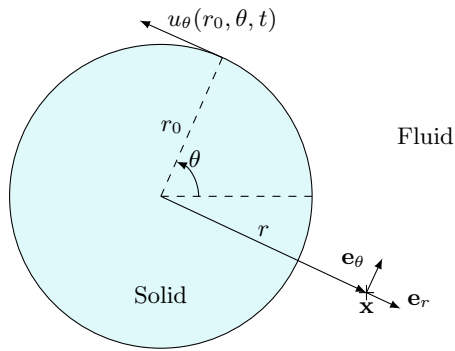

The general solution obtained for the oscillating and vibrating cylinder is particularised to the case of the undeformable oscillating cylinder. The direction of rotation of the cylinder alternates sinusoidally with a period $\Omega$ and with a maximal tangential velocity amplitude on the cylinder surface corresponding to $U$. The case considered is depicted in figure \ref{fig:sketchOscillatingCase}.
The problem uses the boundary condition
\begin{equation}
    \begin{array}{rl}
         \vc{u}(r_0,\theta,t) &\hspace{-0.2cm} = U e^{-i\Omega t} \vc{e}_\theta\\
         &\hspace{-0.2cm} = U e^{-i\Omega t} \left( -\sin(\theta) \vc{e}_x + \cos(\theta) \vc{e}_y \right) 
         \ .
    \end{array}
\end{equation}

\subsection{Computation of the velocity field}

The solution is sought in complex form, and the physical meaningful velocity corresponds to the real part of the final result.
By making the different azimuthal mode order of the general solution derived in {\S} \ref{txt:generalSolutionCylinder} correspond,
\begin{equation}
    \begin{array}{ll}
    \displaystyle 
    \dfrac{\sigma_0 + \tau_0}{r_0}
    = (\alpha_0 + \beta_0) e^{-\nu \lambda t} \int_{r_0}^\infty
    \left(\dfrac{r_0}{r_s}\right)^{-1} H_0^{(1)}(\sqrt{\lambda}r_s) 
    \df r_s
    &\text{(far field)} 
    \ ,\\[2ex]
    \displaystyle 
    \dfrac{\sigma_0 + \tau_0}{r_0}
    = - U e^{-i\Omega t}
    &\text{(azimuthal velocity)} 
    \ ,\\[2ex]
    \displaystyle 
    \sigma_n
    = \dfrac{r_0}{2n} \alpha_n e^{ - \nu \lambda t}
    \int_{r_0}^\infty
    \left(\dfrac{r_0}{r_s} \right)^{n-1} H_n^{(1)}(\sqrt{\lambda}r_s)
    \df r_s
    &\text{(azimuthal velocity)} 
    \ ,\\[2ex]
    \displaystyle 
    \tau_n
    = \dfrac{r_0}{2n} \beta_{n} e^{ - \nu \lambda t}
    \int_{r_0}^\infty
    \left(\dfrac{r_0}{r_s} \right)^{n-1} H_n^{(1)}(\sqrt{\lambda}r_s)
    \df r_s
    &\text{(azimuthal velocity)} 
    \ ,\\[2ex]
    \displaystyle 
    \sigma_n  
    = - \dfrac{r_0}{2n}
    \alpha_n e^{ - \nu \lambda t}
    \int_{r_0}^\infty
    \left(\dfrac{r_0}{r_s} \right)^{n-1} H_n^{(1)}(\sqrt{\lambda}r_s)
    \df r_s
    &\text{(radial velocity)}  
    \ ,\\[2ex]
    \displaystyle 
    \tau_n 
    = - \dfrac{r_0}{2n}
    \beta_{n} e^{ - \nu \lambda t}
    \int_{r_0}^\infty
    \left(\dfrac{r_0}{r_s} \right)^{n-1} H_n^{(1)}(\sqrt{\lambda}r_s)
    \df r_s
    &\text{(radial velocity)}  
    \  ,\\[2ex]
    \end{array}
\label{eq:matching_BC_undeformable_cylinder}
\end{equation}
with $n\in\mathbb{N}^*$. For $n\neq0$, $\sigma_n=\tau_n=0$, leading either to the trivial solution $\alpha_n=\beta_n=0$ or to the generation of flow eigenmodes obtained for $\lambda$ such as,
\begin{equation}
    \int_{r_0}^\infty
    \left(\dfrac{r_0}{r_s} \right)^{n-1} H_n^{(1)}(\sqrt{\lambda}r_s)
    \df r_s = 0
    \hspace{0.5cm}
    \text{with Re$(\lambda)\ge0$ and Im$(\lambda)\ge0$.}
\end{equation}
The existence of these eigenmodes is possible according to this linear theory, but the amplitudes of the modes, if any, have to be determnined by a non-linear model. These solution are discarded here, and only the zero-th mode order is considered in this problem. The matching for this lowest mode provides,
\begin{equation}
    (\alpha_0 + \beta_0) \int_{r_0}^\infty
    \left(\dfrac{r_0}{r_s}\right)^{-1} H_0^{(1)}(\sqrt{\lambda}r_s) 
    \df r_s = - U e^{(\nu \lambda -i\Omega) t}
    \ .
\end{equation}
The coefficients $\alpha_0$ and $\beta_0$ are independent of time, so is the complete left-hand side of the previous equation. For the right-hand side of the previous equation similarly to be time independent, $\lambda$ can only take the value,
\begin{equation}
    \lambda = \dfrac{i\Omega}{\nu}
    \ .
\end{equation}
Using the definition of the streamfunction with $\psi = \psi_G+\psi_H$ leads to
\begin{equation}
    u_r(r,\theta,t) = 0
    \hspace{0.5cm} \text{and} \hspace{0.5cm}
   u_\theta(r,\theta,t) = \mathfrak{Re} \left(
   \dfrac{\displaystyle \int_{r}^{\infty} \left(\dfrac{r}{r_s}\right)^{-1} H_0^{(1)}\left(\sqrt{\dfrac{i\Omega}{\nu}}r_s\right) \df r_s}{\displaystyle \int_{r_0}^{\infty} \left(\dfrac{r_0}{r_s}\right)^{-1} H_0^{(1)}\left(\sqrt{\dfrac{i\Omega}{\nu}}r_s\right) \df r_s} U e^{-i\Omega t}
   \right)
   \label{eq:veloSol_OscilCyl}
\end{equation}
where the real part operator $\mathfrak{Re}()$ has to be applied on the solution to be physically meaningful. 
It is shown in Appendix \ref{txt:equivalenceWithMorfey}, that this solution in fact corresponds to the solution given in previous studies \citep{Meecham_jasa65, Lauvstad_jasa68, LauvstadMeecham_jsv69, Meecham_jsp73, MorfeySorokinWright_jsv22}, that is,
\begin{equation}
    u_r(r,\theta,t) = 0
    \hspace{0.5cm} \text{and} \hspace{0.5cm}
    u_\theta(r,\theta,t) = \mathfrak{Re} \left( 
    \dfrac{H_1^{(1)}\left(\sqrt{\dfrac{i\Omega}{\nu}}r\right)}{H_1^{(1)}\left(\sqrt{\dfrac{i\Omega}{\nu}}r_0\right)}  U e^{-i\Omega t}
    \right)
\end{equation}
Denoting by $\overline{z}$ the complex conjugate of $z$, the following relation holds: $\overline{H_n^{(1)}(z)} = H_n^{(2)}(\overline{z})$. Since $\overline{\sqrt{i}} = -i\sqrt{i}$
and introducing the modified Bessel function of second kind $K_n(x) = \frac{\pi}{2}(-i)^{n+1}H_n^{(2)}(-ix)$ the previous expression becomes,
\begin{equation}
    u_r(r,\theta,t) = 0
    \hspace{0.5cm} \text{and} \hspace{0.5cm}
    u_\theta(r,\theta,t) = 
    \dfrac{U}{2} \left(
    \dfrac{H_1^{(1)}\left(k_\nu r\right)}{H_1^{(1)}\left(k_\nu r_0\right)}e^{-i\Omega t}
    + \dfrac{K_1\left(k_\nu r\right)}{K_1\left(k_\nu r_0\right)}e^{i\Omega t}
    \right)
    \label{eq:solution_Morfey}
\end{equation}
where $k_\nu=\sqrt{i\Omega/\nu}$ is introduced.
It appears that the velocity solution of this incompressible flow problem given in equations (\ref{eq:solution_Morfey}) perfectly agrees with the velocity field solution of the linearised compressible Navier-Stokes equations. The solution to this compressible flow problem was first derived by \citet[App. D]{MorfeySorokinGabard_jfm12}\footnote{It seems that (D12) and (D13) of this reference are solely valid for $m=0$. Indeed $V_m=1$ holds only for this mode order. Note also the slight difference between the result given here for mode zero and \cite[eq. (10)]{MorfeySorokinWright_jsv22}.}, and reviewed in Appendix \ref{txt:solutionToTheCompressibleProblem} of the present study.
At this level of approximation, the solution to the compressible Navier-Stokes equations is fully isobar and no sound is radiated.
The next section illustrates how non-linearities can be calculated in the computation of the pressure field, and the acoustic field calculated.

\subsection{Computation of the pressure field}

The pressure field of incompressible flows is an instantaneous imprint of the velocity field as defined by Poisson's equation (\ref{eq:Poisson's_eq}). Considering the dependencies in $r$ and $\theta$ of $u_r$ and $u_\theta$ given in equation (\ref{eq:solution_Morfey}), it follows that\footnote{Where for an incompressible flow, $\nabla \cdot \nabla \cdot (\rho \vc{u} \otimes \vc{u}) = \rho \text{tr}\left( (\nabla \vc{u})(\nabla \vc{u})\right)$. The right-hand side expression is then obtained recalling that in a polar coordinate system $\nabla \vc{u} = \partial u_r / \partial r \, \vc{e}_r \otimes \vc{e}_r 
+ \left( (1/r) \partial u_r / \partial \theta - u_\theta/r \right) \, \vc{e}_r \otimes \vc{e}_\theta 
+ \partial u_\theta / \partial r \, \vc{e}_\theta \otimes \vc{e}_r 
+ \left( (1/r) \partial u_\theta / \partial \theta + u_r/r \right) \, \vc{e}_\theta \otimes \vc{e}_\theta$.} 
\begin{equation}
    \Delta p_i =  \dfrac{1}{r} \dfrac{\partial ( \rho_i u_\theta^2(r))}{\partial r} H(r-r_0)
    \hspace{1cm} \text{and} \hspace{1cm}
    \left\{
    \begin{array}{l}
         \left. \dfrac{\partial p_i}{\partial r} \right|_{r=r_0,t}= \dfrac{\rho_i u_\theta^2(r_0,t)}{r_0} 
         \ ,\\
         \displaystyle \lim_{r\rightarrow\infty} \left(p_i\right) = 0
         \ ,
    \end{array} \right.
    \label{eq:PoissonEq2RebuiltPress}
\end{equation}
where the Neumann boundary condition in the direction normal to the cylinder's surface comes from the projection of the momentum equation along the radial direction $\vc{e}_r$.
This problem is very similar to the one tackled in {\S} \ref{txt:takingAcountBC}, and, as previously, the solution is found as a composition of an inhomogeneous free-field solution $p_G$ together with a homogeneous solution of Poisson's equation $p_H$,
\begin{equation}
    p_i = p_G + p_H 
    \ .
\end{equation}
The application of Green's formula with Green's function (\ref{eq:Poisson's_Green's_fct}) leads, after an integration by part, to
\begin{equation}
    p_G(r,t) = - \ln\left( \dfrac{r}{r_0}\right) \rho_i u_\theta^2(r_0,t)
    - \int_{r}^\infty \dfrac{\rho_i u_\theta^2(r_s,t)}{r_s} \df r_s
    \ .
\end{equation}
To guarantee the far-field boundary condition, the following particular eigensolution of the problem must be chosen (compare with equation (\ref{eq:homogeneous_solution_psi_H})),
\begin{equation}
    p_H(r,t) = \ln\left( \dfrac{r}{r_0}\right) \rho_i u_\theta^2(r_0,t)
    \ ,
\end{equation}
so that 
\begin{equation}
    p_i(r,t) = - \int_{r}^\infty \dfrac{\rho_i u_\theta^2(r_s,t)}{r_s} \df r_s
    \ ,
    \label{eq:pressureEvolution}
\end{equation}
where $u_\theta(r,t)$ corresponds to the solution given in equation (\ref{eq:solution_Morfey}). 
Note, that the previous solution automatically fulfils the boundary condition on the cylinder surface.
It is moreover found that this solution is in agreement with the expression of \citet{Meecham_jasa65, Meecham_jsp73} which, so far, remained unproved.
Analytical solution of integrals involving Bessel functions very similar to the one arising in the previous expression exist in the literature \cite[\href{https://dlmf.nist.gov/10.22.7}{(10.22.7)}]{NIST_DLMF_web23}
, but from these authors' attempts, those do not seem to work when modified Bessel function of the second kind $K_n$ are involved.
Therefore solutions to equation (\ref{eq:pressureEvolution}) will be approximated in what follows by numerical integration.

Before using the knowledge of the incompressible velocity and pressure fields to provide an exact solution to the acoustic radiation problem, the validity of these expressions will be verified numerically.

\subsection{Numerical verification of the solution}
\label{txt:numericalMethod}

Dimensionless variables associated with this configuration are first introduced.
This problem possesses the four following independent parameters $r_0$, $U$, $\Omega$ and $\nu$ which are combinations of time and distance. From the $\Pi$-theorem, two dimensionless parameters can be built.
The Stokes number $S$ and the Reynolds number Re are chosen as the main parameters to study this problem. They are introduced as,
\begin{equation}
S = \dfrac{r_0^2 \Omega}{\nu}
\hspace{0.5cm} \text{,} \hspace{0.5cm}
\text{Re} = \dfrac{r_0 U}{\nu}
\end{equation}
The scaled angular frequency $\Theta = \nu \Omega/U^2$ introduced in \citet{MorfeySorokinWright_jsv22}, or the kinematic number that compares the frequency at which the cylinder changes rotation with respect to its angular speed of rotation $\Xi = r_0 \Omega/U$ could alternatively be chosen. Any of these two dimensionless numbers can be constructed from the others, e.g. $\Theta=S/\text{Re}^2$ and $\Xi = S/\text{Re}$. 

The analytical solution derived here for the velocity field assumes linearity and thus that the advection term $(\vc{u} \cdot \nabla) \vc{u}$ plays a lesser role in the momentum equation than the viscous term $\nu \Delta \vc{u}$, and has been neglected in the derivations, see equation (\ref{eq:homogeneous_vorticity_eq_NABLA}). The dimensionless number that compares these quantities is the Reynolds number.
It is expected that the analytical solution introduced here is acceptable at low Reynolds number. The validity of this approximation is assessed numerically in what follows.

The implementation of the incompressible Navier-Stokes equations by \citet{FrancoCamierAndrejPazner_cf20} based on the finite element library \texttt{MFEM} \citep{mfem_general_cite_cma21} developed at Lawrence Livermore National Laboratory (LLNL) is chosen for this work.
This flow problem is solved using a high-order finite element discretization in space and a third order implicit-explicit method in time that leverages an extrapolation scheme for the convective parts and a backward-difference formulation for the viscous parts of the equation.
An implementation of the spatial filter of \citet{FischerMullen_crasp01} is used to stabilise the calculation of turbulent flows and filter out the scales of the flow that cannot be resolved by the grid size and order of the elements considered.
A similar technique is used by \citet{BogeyBailly_cf06} to perform Large Eddy Simulation (LES) of jets without a subgrid model.
Numerical truncation of the computational domain is achieved using a buffer zone in which a spring force is applied in the momentum equation to gradually restore $\vc{u}_0$, the incompressible velocity field, to the ambient condition. The spring force strength follows a cubic evolution law. The outer boundaries are then simply enforced with Dirichlet boundary conditions.

\begin{figure}
    \centering
    \begin{tikzpicture}
        \node at (0cm,0cm) {\includegraphics[height=4.5cm, trim=0.0cm 0.0cm 0.0cm 0.0cm,clip]{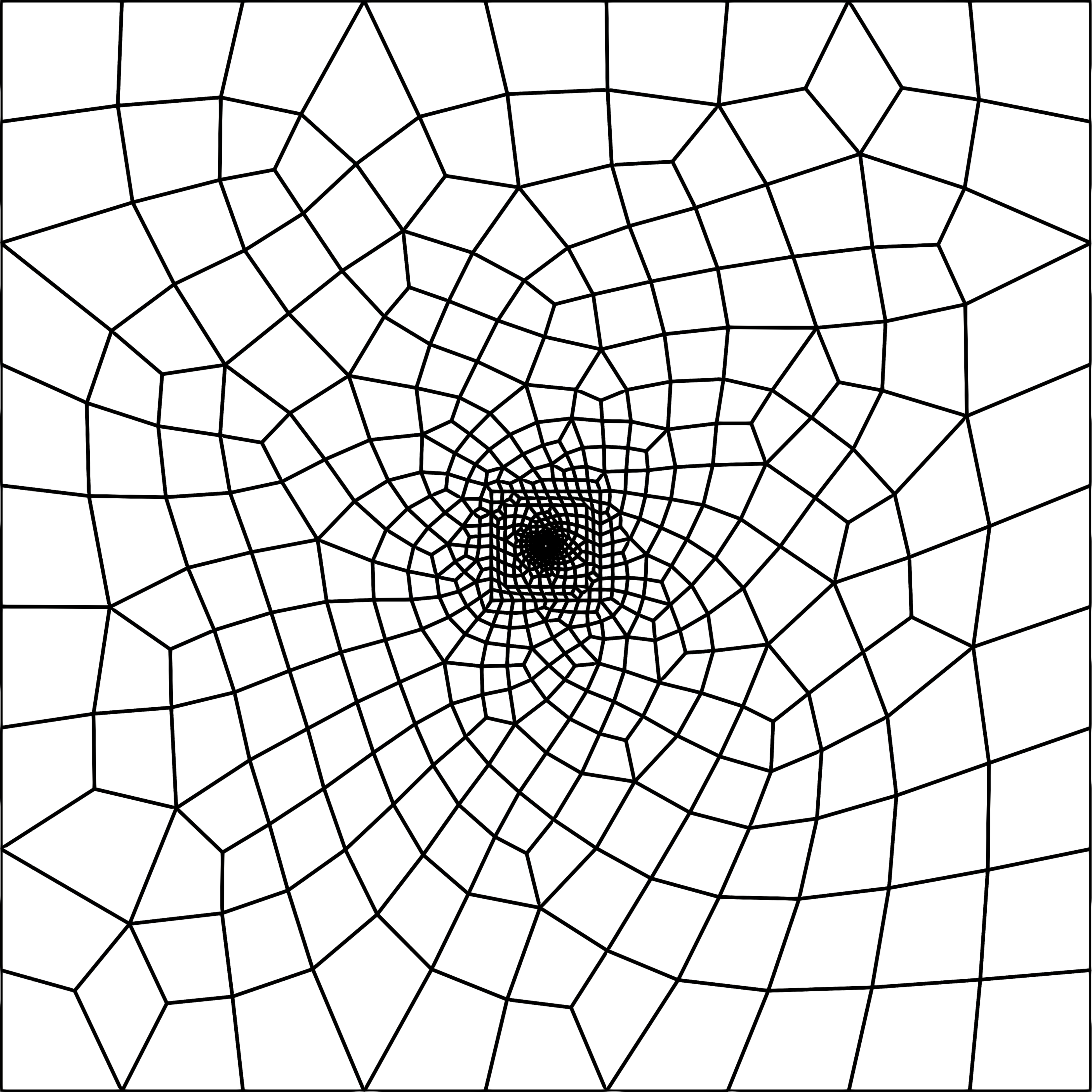}};
        \node at (6.5cm,0cm) {\includegraphics[height=4.5cm, trim=0.0cm 0.0cm 0.0cm 0.0cm,clip]{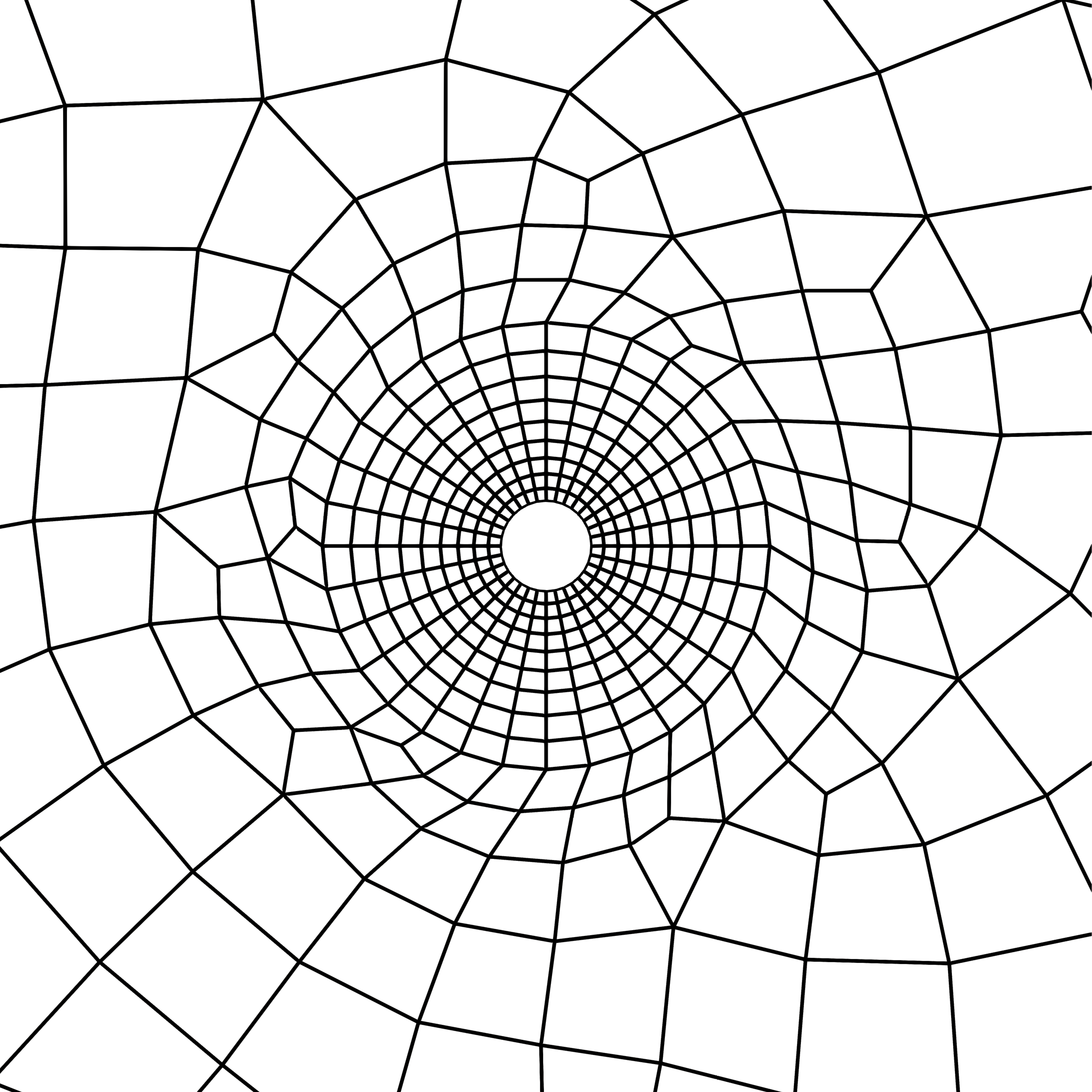}};
    \end{tikzpicture}
    \caption{Extent of the total computation domain (left), and close-up of the structured layer around the cylinder (right).}
    \label{fig:NumericalDomain}
\end{figure}

The open source mesh generator \texttt{Gmsh} \citep{GeuzaineRemacle_ijnme09} is used to discretise the problem. A customisable script to create partially structured grids of quadrilateral curved elements is created. Elements with order 8 are considered to represent the solution. The grid used to compute the first configuration studied in this work is presented in figure \ref{fig:NumericalDomain}. 
If the radius of the cylinder is $r_0$, the physical domain is a square with $60 r_0$ long edges centred around the cylinder. The sponge zone surrounds the physical domain so that the total computational domain in a square with $800 r_0$ long edges centred around the cylinder. 
$32$ elements are used azimuthally in the structured grid that forms a layer around the cylinder. The latter is composed of $9$ layers of elements that stretch over a width of $4r_0$, that is from $r_0$ to $5r_0$ with a geometric progression of the element size of $1.1$.
This problem is bidimensional and the numerical cost associated with the simulations performed here are negligible. Finer grids are thus used when needed to ensure a resolution of all scales of the flow. The mesh topology and domain size remains identical in each case.

\begin{figure}
    \centerline{
    \includegraphics[scale=0.92 ,trim=0.2cm 0.0cm 0.2cm 0.0cm,clip]{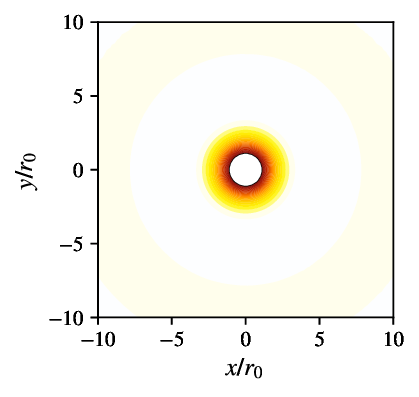}
    \includegraphics[scale=0.92 ,trim=0.2cm 0.0cm 0.0cm 0.0cm,clip]{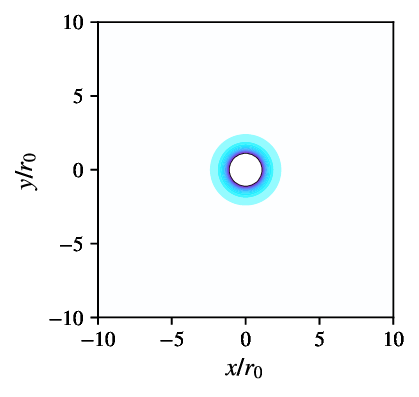}
    }
    \caption{Contour of the tangential velocity $u_\theta$ (left) and the incompressible pressure field $p_i$ (right) solved with \texttt{MFEM} for $S=1.0$ and Re$=1.0$ at the normalised time $t\Omega = 100\pi$.}
    \label{fig:cartoS1Re1}
\end{figure}

Figure \ref{fig:cartoS1Re1} presents the solution for the tangential velocity $u_\theta$ and the incompressible pressure $p_i$ obtained for this problem for Stokes and Reynolds numbers equal to $1.0$. The problem is clearly axisymmetric and higher-order eigensolutions of the problem mentioned in {\S} \ref{txt:undeformableCylinder} are not contributing to the solution.
The solution is presented for the time $t\Omega=100\pi$ which corresponds to the time for which the tangential velocity of the cylinder is maximal. At this moment, the amplitude of the velocity field and of the pressure field both are maximum at the cylinder surface and decay rapidly with increasing radial distance.
The maximal value of the normalised radial velocity $|u_r|/U$ is less than $10^{-8}$, so it is taken to be zero and discarded from the discussions.

\begin{figure}
    \centerline{
    \includegraphics[scale=0.92 ,trim=0.2cm 0.0cm 0.2cm 0.0cm,clip]{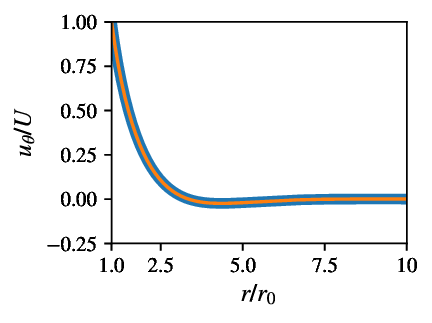}
    \includegraphics[scale=0.92 ,trim=0.2cm 0.0cm 0.0cm 0.0cm,clip]{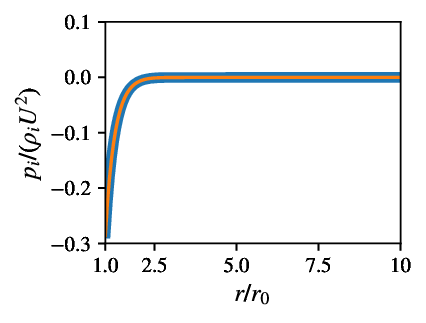}
    }
    \caption{Comparison of the incompressible flow solution for $S=1.0$ and Re$=1.0$ at $t\Omega=100\pi$. Radial evolution of, the tangential velocity $u_\theta$ (left), and, the pressure $p_i$ (right); ({\color[RGB]{31,119,180}{\fullfat}}) solution computed with \texttt{MFEM}, and ({\color[RGB]{255, 127, 14}{\full}}) equations (\ref{eq:solution_Morfey}) and (\ref{eq:pressureEvolution}) respectively.}
    \label{fig:cartoS1Re1_slice}
\end{figure}

\begin{figure}
    \centerline{\includegraphics[scale=0.92 ,trim=0.2cm 0.0cm 0.2cm 0.0cm,clip]{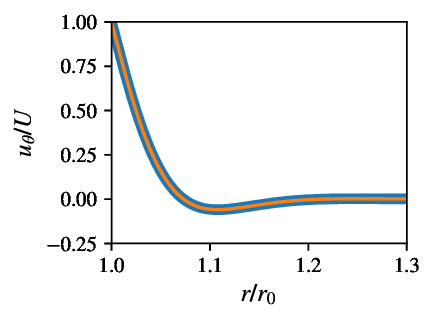}
    \includegraphics[scale=0.92 ,trim=0.2cm 0.0cm 0.2cm 0.0cm,clip]{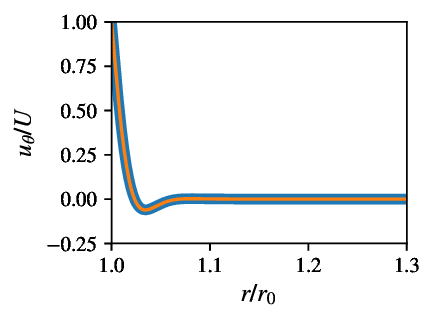}}
    \caption{Comparison of the tangential velocity $u_\theta$ at $t\Omega=10\pi$, for $S=10^3$ and Re$=1$ (left), and $S=10^4$ and Re$=10^2$ (right);
    ({\color[RGB]{31,119,180}{\fullfat}}) solution computed with \texttt{MFEM}, ({\color[RGB]{255, 127, 14}{\full}}) equation (\ref{eq:solution_Morfey}).}
    \label{fig:cartoVariousSandRe_slice}
\end{figure}

The result of the numerical simulation is compared in figure \ref{fig:cartoS1Re1_slice} with the analytical solution developed here on a radial sample. Both the velocity and the pressure fields simulated match perfectly the analytical solution derived.
Results for the tangential velocity $u_\theta$ obtained with the incompressible Navier-Stokes solver are compared in figure \ref{fig:cartoVariousSandRe_slice} to the output given by the linear model for various arbitrary chosen values of the Stokes and the Reynolds number. 
As for the results presented in figure \ref{fig:cartoS1Re1_slice}, results shown in figure \ref{fig:cartoVariousSandRe_slice} indicate a seemingly perfect agreement between the computed solution and the theory.
In particular, no significant effect of non-linearities is observed for the pair of parameters tested. These non-linearities would yet intuitively been expected as the Reynolds number increases. This effect of the Reynolds number is now investiguated.

\begin{figure}
    \centerline{\includegraphics[scale=0.92 ,trim=0.2cm 0.0cm 0.2cm 0.0cm,clip]{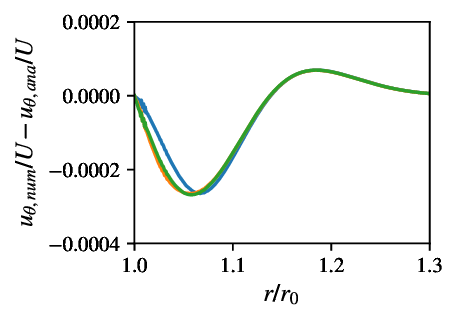}}
    \caption{Deviation with respect to the analytical solution for various Reynolds numbers Re, for $S=10^4$ at $t\Omega = 10 \pi$; ({\color[RGB]{31,119,180}{\full}}) Re$=10^2$, ({\color[RGB]{255, 127, 14}{\full}}) Re=$10^3$ and ({\color[RGB]{44, 160, 44}{\full}}) Re=$10^4$.}
    \label{fig:ReynoldsIndependanceStudy}
\end{figure}

From the analytical solution provided in equation (\ref{eq:veloSol_OscilCyl}),
\begin{equation}
    \dfrac{u_\theta}{U} \propto \dfrac{H_1^{(1)}\left(\sqrt{iS}\ r/r_0 \right)}{H_1^{(1)}\left( \sqrt{iS} \right)}
\end{equation}
the normalised tangential velocity solely depends on the Stokes number and appears to be independent of the Reynolds number.
A parametric study is conducted to assess the effect of the Reynolds number on the non-linear solution computed with the incompressible flow solver to estimate the approximation of this linear solution. The Stokes number is set to $S = 10^4$ while the Reynolds number takes the value $10^2$, $10^3$ and $10^4$. An identical grid, similar but finer to the one shown in figure \ref{fig:NumericalDomain}, is used for this study.
Table \ref{tab:configStudied} summarises the cases studied here, and indicates the correspondence with the other dimensionless variables. The cases (c), (d) and (e) are considered for this study.
\begin{table}
\begin{center}
\begin{tabular}{ c|cccc } 
 & Re & $S\ $ & $\Theta$ & $\Xi$ \\ 
 \hline
 case (a) & $1$ &  $1$ & $1$ & $1$ \\[-1.5ex] 
 case (b) & $1$ & $10^3$ & $10^3$  & $10^3$\\[-1.5ex] 
 case (c) & $10^2$ & $10^4$ & $1$ & $10^2$\\[-1.5ex]
 case (d) & $10^3$ & $10^4$ & $10^{-2}$ & $10$\\[-1.5ex]
 case (e) & $10^4$ & $10^4$ & $10^{-4}$ & $1$\\
\end{tabular}
\label{tab:configStudied}
\caption{Table of dimensionless numbers summarising the simulations carried out.}
\end{center}
\end{table}
The deviation of the tangential velocity $u_\theta$ computed with the non-linear flow solver is compared to the analytical solution given in equation (\ref{eq:solution_Morfey}) and presented in figure \ref{fig:ReynoldsIndependanceStudy}. Solutions are computed and compared a time $t\Omega = 10\pi$. The results are presented on a radial sample extending up to $1.3r_0$. Deviations between analytical and numerical simulations become indeed smaller by several order of magnitudes for larger radius.
The dimensionless deviations with respect to the analytical solution are as small as $3\times10^{-4}$, and even if three decades of Reynolds number are considered, the latter deviations do not vary significantly. Finally, it appears that the critical Reynolds number for this configuration is very high (higher than $10^4$) and that the range of validity of the analytical solution derived here seems valid up to the point where the flow transitions to turbulence.

From the $\Pi$-theorem evoked earlier, two independent dimensionless parameters are sufficient to describe the physics under study. Since the solution of this problem is Reynolds independent, a single parameter, such as the Stokes number, is sufficient to characterise the solution.
Since the phase of the solutions presented in figure \ref{fig:cartoS1Re1_slice} and \ref{fig:cartoVariousSandRe_slice} are identical, the modifications of the flow profiles in these configurations are hence directly attributed to the variation of the Stokes number.


\section{Calculation of the acoustic far-field response at low Mach number}
\label{txt:acousticFarFieldSol}

The analytical expressions of the incompressible velocity and pressure fields of the problem under consideration has been established and validated. They are now used to derive the acoustic far-field response. 
In section \ref{txt:failureOfLighthill} some specificity of sound generation at low Mach numbers are recalled. After highlighting that, in the limit of the low Mach numbers, Lighthill's acoustic analogy (with the source approximation $\vc{T} \approx \rho_i \vc{u}_i \otimes \vc{u}_i$) coincides with Powell's and Ribner's acoustic analogy, this section applies these theories to the case of the undeformable oscillating cylinder tackled in {\S} \ref{txt:undeformableCylinder}.

\subsection{Using the asymptotic expression of Powell's vortex sound theory}
\label{txt:Powell_Ac_Analogy}

\citet[eq. (25)]{Powell_jasa64} proposed the following asymptotic expression for low-speed flows,
\begin{equation}
    \dfrac{\partial^2 p}{\partial t^2} - a_0^2 \Delta p = a_0^2 \nabla \cdot \left( \rho_i \left(\nabla \times \vc{u}_i \right) \times \vc{u}_i + \dfrac{\nabla(\rho_i \vc{u}_i)^2}{2}  \right)
    \ .
\end{equation}
Provided that previous right-hand side does not vanish, this approximation is exact to the second order in Mach number.
It is straightforward\footnote{Using the formulas $(\nabla \times \vc{u}) \times \vc{v} = (\nabla \vc{u}) \cdot \vc{v} -(\nabla \vc{u})^T \cdot \vc{v}$ and $\nabla(\vc{u} \cdot \vc{v}) = (\nabla \vc{u})^T \cdot \vc{v} + (\nabla \vc{v})^T \cdot \vc{u}$, together with $\nabla \cdot (\vc{T} \cdot \vc{u}) = (\nabla \cdot (\vc{T}^T))\cdot \vc{u} + \text{tr}(\vc{T}\cdot (\nabla \vc{u}))$ and previous calculus relations already recalled.} to show that this right-hand side for an incompressible flow field corresponds to $\rho_i a_0^2 \text{tr}((\nabla \vc{u}_i) (\nabla \vc{u}_i))$ which is indeed consistent with the incompressible source term of Lighthill's given in equation (\ref{eq:Lighthill_incompSource}).
Finally by using Poisson's equation for incompressible flow, given in equation (\ref{eq:Poisson's_eq}), this terms can be recast to obtain
\begin{equation}
    \dfrac{\partial^2 p}{\partial t^2} - a_0^2 \Delta p = - a_0^2 \Delta p_i
    \ .
    \label{eq:Powell-Ribner}
\end{equation}
This asymptotic formulation of Powell's acoustic analogy does not degenerate at low Mach numbers as in the naive implementation of Lighthill's acoustic analogy discussed in {\S} \ref{txt:failureOfLighthill}.
In his work, \cite{Powell_jasa64} did not split the pressure field $p$, that appears in his wave operator, into a hydrodynamic and an acoustic part. From equation (\ref{eq:Powell-Ribner}), the reader clearly sees that, if this split is done, this formulation becomes identical to Ribner's dilatation theory used hereafter to compute the sound radiated.

\subsection{Using Ribner's dilatation theory}
\label{txt:RibnersDilatationTheory}

In his pioneer work, \citet{Ribner_utia62} postulated the split of the fluid pressure $p$ into an acoustic dilatation $p'$ part, and an incompressible hydrodynamic part $p_i$, the so-called pseudo-sound, such as $p=p'+p_i$. Using these notations, Ribner's dilatation theory is
\begin{equation}
    \dfrac{\partial^2 p'}{\partial t^2}  - a_0^2 \Delta p' = - \dfrac{\partial^2 p_i}{\partial t^2}
    \ .
    \label{eq:RibnersDilationTheory}
\end{equation}
Expansion about incompressible flow \citep{HardinPope_tcfd94, ShenSorensen_aiaaj99, ShenSorensen_aiaaj01, SlimonSoteriouDavis_jcp00}
or more advanced theories \citep{SeoMoon_jcp06, MunzDumbserRoller_jcp07, MoonSeoBaeRogerBecker_cf10, KaltenbacherHuppeReppenhagenTautzBeckerKuehnel_sae16}, all correspond at the leading order to Ribner's model.

\begin{figure}
    \centerline{\includegraphics[scale=0.92 ,trim=0.2cm 0.0cm 0.2cm 0.0cm,clip]{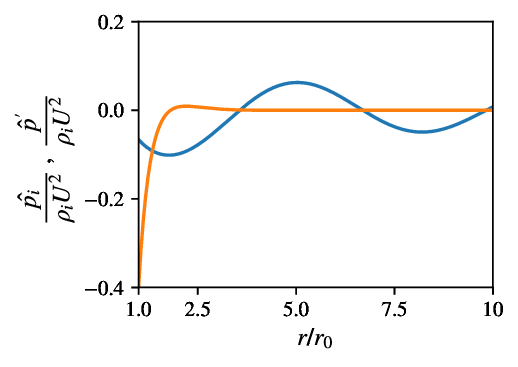}}
    \caption{Normalised amplitudes of ({\color[RGB]{31,119,180}{\full}}) the acoustic pressure, and ({\color[RGB]{255, 127, 14}{\full}}) the hydrodynamic pressure. The pair $S=1.0$ and $K=1.0$ is considered for the frequency $\omega = 2\Omega$.}
    \label{fig:acousticVsHydroK1S1Re1}
\end{figure}

Since tailored Green's functions that account for the scattering of sound by cylinders, only exist in the frequency domain, the sound radiation problem is solved considering a Fourier transformation\footnote{By convenience, in what follows $\omega$ defines an angular frequency associated with the Fourier analysis, and not any more the norm of the vorticity.}.
The convention chosen for the Fourier analysis is defined as $\hat{\phi}(r,\omega) = \mathcal{F}(\phi(r,t)) = \int_{-\infty}^{\infty} \phi(r,t) e^{i\omega t} \df t$, so that the above equation becomes, in the frequency domain,
\begin{equation}
    (k^2 + \Delta) \hat{p}' = -k^2 \hat{p_i}
    \label{eq:Helmholtz_eq}
\end{equation}
where $k=\omega/a_0$ and $\hat{p_i}(r,\omega) = \mathcal{F}(p_i(r,t)) $. Considering equations (\ref{eq:solution_Morfey}) and (\ref{eq:pressureEvolution}), and retaining only acoustic meaningful frequencies, i.e. $\omega>0$,
\begin{equation}
\hat{p_i}(r,\omega) = - \pi \delta(\omega - 2\Omega) \dfrac{\rho_i U^2}{2} \int_r^\infty \dfrac{1}{r_\alpha}
\dfrac{H_1^{(1)}\left( k_\nu  r_\alpha \right)^2}{H_1^{(1)}\left( k_\nu  r_0 \right)^2} \df r_\alpha    
\end{equation}
This time this integral can be computed following \cite[\href{https://dlmf.nist.gov/10.22.7}{(10.22.7)}]{NIST_DLMF_web23}
, and one obtains
\begin{equation}
\hat{p_i}(r,\omega) = - \pi \delta(\omega - 2\Omega) \dfrac{\rho_i U^2}{4} 
\dfrac{H_1^{(1)}\left( k_\nu  r \right)^2 + H_0^{(1)}\left( k_\nu r \right)^2}{H_1^{(1)}\left( k_\nu  r_0 \right)^2} 
\ .
\end{equation}
Green's function for the acoustic radiation problem must verify Helmholtz's equation.
Since the sound source is purely axisymmetric, the $\theta$-dependence in the wave operator is discarded, and Green's function is the solution of
\begin{equation}
    \left( k^2 + \dfrac{1}{r} \dfrac{\partial}{\partial r} \left( r \dfrac{\partial }{\partial r}\right) \right) G_{r_s}(r,\omega)= \dfrac{1}{2\pi r} \delta (r-r_s)
    \ .
    \label{eq:Helmholtz_Green_eq}
\end{equation}
In principle, any boundary condition for previous Green's function can be chosen. The effect of the scattering by the cylinder simply has to be consistently accounted for in the integral formulation \citep{GloerfeltPerotBaillyJuve_jsv05}. Previous studies \citep{Meecham_jasa65, LauvstadMeecham_jsv69} have highlighted that the volume source term and the surface source term are of comparable order. To consider only the volume source contribution, Green's function tailored to the geometry is chosen, and the following Neumann boundary conditions are considered: $\partial G_{r_s}(r,\omega)/\partial r = 0$ at $r=r_0$, and $\partial G_{r_s}(r,\omega)/\partial r_s= 0$ at $r_s=r_0$.
The solution for this Green's function can be found in \citet{GloerfeltPerotBaillyJuve_jsv05} and after assuming that the observer is set in the far field\footnote{A similar tailored Green's function but for the first azimuthal mode order of is considered in \citet[{\S} 3]{MorfeySorokinWright_jsv22} in the context of a matched asymptotic solution. In \citet[{\S} 5.4]{MorfeySorokinWright_jsv22} the zero-th mode order is considered in application with a thermoviscous acoustic analogy.}. It follows that
\begin{equation}
    G_{r_s}(r,\omega) = \dfrac{1}{4} J_1(k r_0) Y_1(k r_0) \dfrac{H_0^{(1)}(k r)}{H_1^{(1)}(k r_0)}
    \left( \dfrac{J_0(k r_s)}{J_1(k r_0)} - \dfrac{Y_0(k r_s)}{Y_1(k r_0)} \right)
    \ .\label{eq:acousticTailoredGreensFunction}
\end{equation}
For a more general expression of Green's functions that accounts for higher azimuthal modes, and that is additionally valid in the acoustic near-field, the reader can refer to \citet{GloerfeltPerotBaillyJuve_jsv05}.
Finally, the acoustic far-field response in the frequency domains for the present configuration becomes,
\begin{equation}
    \hat{p}'(r,\omega) = -k^2 \int_{r_0}^\infty \int_{0}^{2\pi} G_{r_s}(r,\omega) \hat{p_i}(r_s,\omega) r_s \df \theta \df r_s
\end{equation}
That is,
\begin{equation}
    \hat{p}'(r,\omega) = \alpha
    \dfrac{H_0^{(1)}(k r)}{H_1^{(1)}(k r_0)}    
    \delta(\omega - 2\Omega)
    \int_{r_0}^\infty 
    \chi(r_s)
    \xi(r_s)
    r_s \df r_s
    \label{eq:acousticSolWithCylinder}
\end{equation}
with $\chi(r) = ( J_0(k r)/J_1(k r_0) - Y_0(k r)/Y_1(k r_0) )$ and
$\xi(r) = H_1^{(1)}( k_\nu  r )^2 + H_0^{(1)}( k_\nu r )^2 $, together with
$\alpha =  -k^2 \pi^2 \rho_i U^2 J_1(k r_0) Y_1(k r_0)/(8 H_1^{(1)}\left( k_\nu  r_0 \right)^2)$.
The evolution of the acoustic pressure $\hat{p}'$ and of the hydrodynamic pressure $\hat{p_i}$ are comapred in figure \ref{fig:acousticVsHydroK1S1Re1}. The normalised spectral amplitudes at the frequency $\omega = 2\Omega$ are plotted.
The hydrodynamic solution studied earlier, for which $S=1.0$ and Re$=1.0$, as well as the corresponding acoustic solution for $K=1.0$ are displayed, where the Helmholtz number $K=\omega r_0/a_0$ is a dimensionless measure of the acoustic frequency.

The undeformable oscillating cylinder provides an acoustic response for the parameters considered in figure \ref{fig:acousticVsHydroK1S1Re1}, that is far from negligible.
The pressure field computed within the incompressible flow framework accounts for non-linear contributions stemming from equation (\ref{eq:Poisson's_eq}).
This latter approach is regarded as the more suitable than the fully linear approach for which the problem is found to be isobar, see Appendix \ref{txt:solutionToTheCompressibleProblem} and \cite{MorfeySorokinGabard_jfm12}.

\subsection{Efficiency of the sound generation mechanism}

In the previous paragraph, the acoustic response for a unique pair of Stokes and Helmholtz numbers is provided. What follows discusses in a more systematic way the effect of the Helmholtz number on the acoustic radiation.
As it is known that acoustic compactness increases the acoustic radiation of sound sources in phase \citep{Fuchs_acta79}, the compactness of the cylinder is expected to play a major role in the efficiency of the sound radiation mechanism.

\begin{figure}
    \centerline{\includegraphics[scale=0.92 ,trim=0.2cm 0.0cm 0.2cm 0.0cm,clip]{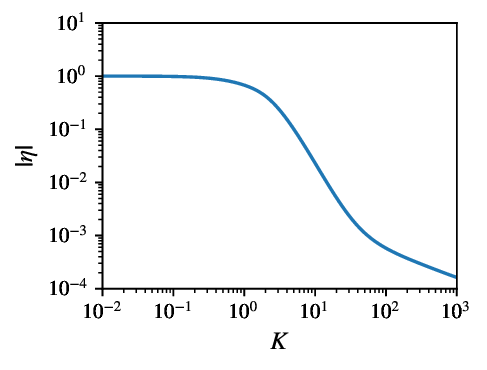}}
    \caption{Evolution of the acoustic efficiency of the oscillating cylinder with $S=1.0$ for different Helmholtz numbers $K$.}
    \label{fig:effectOfCompacity}
\end{figure}

To assess this, the acoustic solution given in equation (\ref{eq:acousticSolWithCylinder}), which accounts for the cylinder scattering effect, is compared to the acoustic solution that would be obtained in absence of the cylinder, with the sound source $-k^2 \hat{p_i}$ lumped together on the cylinder axis.
If $G^{F.F.}$ is free-field Green's function solution of equation (\ref{eq:Helmholtz_Green_eq}), this measure of the acoustic efficiency $\eta$ is
\begin{equation}
    \eta(S,K) = \dfrac{\hat{p}'(r,\omega)}{-k^2 G^{F.F.}_{r_s}(r,\omega) \displaystyle \int_{r_0}^\infty \int_{0}^{2\pi} \hat{p_i}(r,\omega) r_s \df \theta \df r_s }
    \ ,
\end{equation} 
where Green's function $G^{F.F.}$ in two dimensions is \cite[p. 405]{BaillyJuve_aiaaj00,Duffy_15},
\begin{equation}
    G^{F.F.}_{r_s}(r,\omega) = -i\,H_0^{(1)}(kr)/4
    \ .
\end{equation}
The evolution of $|\eta|$ for $S=1.0$ is given in figure \ref{fig:effectOfCompacity}.
For Helmholtz numbers $K$ smaller than one, the acoustic radiation efficiency of the oscillating cylinder is identical to that of a monopole radiating in free field. For larger Helmholtz numbers, the efficiency drops rapidly.
Figure \ref{fig:effectOfCompacity} presents the solution obtained for a particular Stokes number, the dynamic of $|\eta|$ for larger Stokes numbers (not shown) is comparable and presents a rapid drops of the efficiency when $K>1.0$. 
The undeformable oscillating cylinder generates noise when the cylinder is acoustically compact, that is, when the incompressible solution most closely matches the compressible flow dynamics \citep[{\S} 5.6]{GleggDevenport_17}.
Furthermore, the following relation, 
\begin{equation}
    M = \dfrac{\Omega}{\omega} \dfrac{K}{\Xi}
    \label{eq:lowMachNumbercondition}
\end{equation}
links the Helmholtz number and the Mach number, and from equation (\ref{eq:acousticSolWithCylinder}), $\omega = 2\Omega$.
Bearing in mind that acoustic levels decreases with the square of the Mach number (see the coefficient $\alpha$ in equation (\ref{eq:acousticSolWithCylinder})), 
the low Mach number asymptotic limit can realistically be reached for this benchmark case for parameters that are not infinitesimal small. For this to happen, following conditions have to be fulfiled $K \ll \Xi$, i.e. $\text{Re}\ K \ll S$.

\section{Conclusion}

The benchmark case of the flow developing around an oscillating and vibrating cylinder is reviewed.
A careful derivation for the expression of the incompressible pressure field proposed by \citet{Meecham_jasa65} is given, and the incompressible flow solution is derived for non-axisymmetric problems, extending thereby results of the literature.
This work provides the exact expression of the far-field radiated sound by accounting for the cylinder scattering by means of tailored Green's functions.
While an implementation of the general solution presenting azimuthal dependency is exhibited, this work mainly focuses on the undeformable oscillating cylinder configuration.
More complex situation could be tackled by using a resolution strategy identical to the one presented here. For instance, the balance between the flow shear term and the viscous term investigated by \citet{MorfeySorokinWright_jsv22}, omitted in this work, could be investigated for more complex cylinder motions.

A high-order finite element method based on the open source codes \texttt{Gmsh} and \texttt{MFEM} is used to solve the incompressible Navier Stokes equations using curved quadrilateral elements.
The numerical simulations are used to assess the range of validity of the analytical solution. The non-linear advection term is shown to play a negligible role in this configuration, and for all configuration tested, the agreement between the analytical solution and the numerical reference is remarkable.
Both analytical and numerical approach agree that the configuration is independent of the Reynolds number, and the hydrodynamic solution of this configuration is thus solely governed by the Stokes number $S = r_0^2 \Omega/\nu$. The critical Reynolds number leading to a transition to the turbulent regime is found to lie above $10^4$ for this configuration.
Since this flow field is analytically tractable, this benchmark could be a good candidate to study transition mechanisms of boundary layers on curved surfaces.

\section*{Acknowledgements}

The first author has received founding from the Hong Kong Innovation and Technology Commission (ITC): ITS/301/20FP and the Hong Kong Research Grants Council: 16204721.

\section*{Declaration of interests}

The authors report no conflict of interest.

\appendix

\section{Relation between $z H_0^{(1)}(z)$ and $z H_1^{(1)}(z)$}
\label{txt:equivalenceWithMorfey}

The equivalence of both formulation is proven using 
\begin{equation}
    \dfrac{\partial }{\partial z} \left( H_n^{(1)}(z) \right) = \dfrac{n H_n^{(1)}(z)}{z} - H_{n+1}^{(1)}(z)
    \label{eq:recurrence_HK1}
\end{equation}
and,
\begin{equation}
    H_{-n}^{(1)}(z) = (-1)^nH_n^{(1)}(z)
    \label{eq:reflection_HK1}
\end{equation}
after multiplying relation (\ref{eq:recurrence_HK1}) by $z$ and adding $H_n^{(1)}(z)$ to both members of the equality, one obtains,
\begin{equation}
    \dfrac{\partial }{\partial z} \left( z H_n^{(1)}(z) \right) = (n+1) H_n^{(1)}(z) - z H_{n+1}^{(1)}(z)
\end{equation}
Evaluating previous formula for $n=-1$ and using the reflection formula (\ref{eq:reflection_HK1}), one gets,
\begin{equation}
    \dfrac{\partial }{\partial z} \left( z H_1^{(1)}(z) \right) = z H_{0}^{(1)}(z)
\end{equation}
from which the equivalence between equation (\ref{eq:veloSol_OscilCyl}) and equation (\ref{eq:solution_Morfey}) follows.

\section{Implementation of the solution for higher azimuthal order} 
\label{txt:CylinderFlexiSkin}

From the general solution provided in this study, other analytical tractable solution can be deduced. If the cylinder skin is assumed to be flexible, the following boundary condition also fulfils the incompressibility constraints\footnote{To study configurations for which boundary condition do not verify the incompressibility constraint, the compressible equations of Appendix \ref{txt:solutionToTheCompressibleProblem} may be used. If the pressure field computed with this latter approach were to vanish, the non-linear contributions to the pressure field as described by Poisson's equation (\ref{eq:Poisson's_eq}) would be required.}
\begin{equation}
    \begin{array}{rl}
         \vc{u}(r_0,\theta,t) &\hspace{-0.2cm} =  U e^{-i\Omega t} \left( \cos(2\theta) \vc{e}_r - \dfrac{\sin(2\theta)}{2} \vc{e}_\theta \right) \\
         &\hspace{-0.2cm} = U e^{-i\Omega t} \left( \cos^3(\theta) \vc{e}_x - \sin^3(\theta) \vc{e}_y \right)
         \ .
    \end{array}
\end{equation}

This configuration can be studied numerically without mesh deformation in the limit for which the maximal body deformation $\delta_\text{max} = \underset{T,\theta}{\text{max}} ( \int_{T}^{T+\pi/(2\Omega)} u_r(r_0,\theta,t) \df t ) = 2U/\Omega \ll r_0$, that is when $\Xi \gg 1.0$. From equation (\ref{eq:lowMachNumbercondition}), this is also an acceptable scenario to study the low Mach number limit. Indeed, it is then sufficient to attain the low Mach number limit to choose a Helmholtz number so that $K \ll \Xi$.

By matching the boundary conditions for this configuration, as in equation (\ref{eq:matching_BC_undeformable_cylinder}) for the undeformable cylinder, the condition on the azimuthal velocity for the mode order 0 delivers, $\sigma_0 + \tau_0 = 0$, since the far-field condition is unchanged and corresponds to the one given in equation (\ref{eq:matching_BC_undeformable_cylinder}), then $\alpha_0 + \beta_0 = 0$. The solution to the higher mode constraints for $n\neq2$ is also identical to the ones given in equation (\ref{eq:matching_BC_undeformable_cylinder}), hence for $n\neq2$ it follows $\sigma_n = \tau_n = 0$ neglecting the eigenmodes $\alpha_n = \beta_n = 0$. While for $n=2$,
one obtains,
\begin{equation}
    \sigma_2 = -\dfrac{i r_0}{16} U e^{-i\Omega t}
    \hspace{1cm} \text{and} \hspace{1cm}
    \alpha_2 = - \dfrac{3i}{4} \dfrac{U e^{(\nu\lambda-i\Omega) t}}{\displaystyle \int_{r_0}^\infty \left( \dfrac{r_0}{r_s} \right) H_2^{(1)}(\sqrt{\lambda}r_s) \df r_s}
\end{equation}
together with $\tau_2 = -\sigma_2$ and $\beta_2 = - \alpha_2$. Since $\alpha_n$ and $\beta_n$ are time-independent coefficients, $\lambda = i\Omega/\nu$ is retrieved.
Then from the identification with the previous constants, it follows after some calculation,
\begin{equation}
    u_r(r,\theta,t) = \dfrac{3U e^{-i\Omega t}}{4} \cos(2\theta) \left( 
    \dfrac{1}{3}
    \left( \dfrac{r_0}{r} \right)^3 
    + \left(\dfrac{r}{r_0}\right) 
    \left[ 1- 
    \dfrac{
    \displaystyle 
    \int_{r_0}^r \left[ \left( \dfrac{r_0}{r_s}\right) - \dfrac{r_s^3r_0}{r^4} \right] H_2^{(1)}(\sqrt{\lambda}r_s) \df r_s
    }{
    \displaystyle \int_{r_0}^\infty \left( \dfrac{r_0}{r_s} \right) H_2^{(1)}(\sqrt{\lambda}r_s) \df r_s
    }
    \right]
    \right)
\end{equation}
and,
\begin{equation}
    u_\theta(r,\theta,t) 
    = \dfrac{-3U e^{-i\Omega t}}{4} \sin(2\theta) \left(
    - \dfrac{1}{3}
    \left( \dfrac{r_0}{r} \right)^3 
    + \left(\dfrac{r}{r_0}\right) 
    \left[ 1- 
    \dfrac{
    \displaystyle 
    \int_{r_0}^r \left[ \left( \dfrac{r_0}{r_s}\right) + \dfrac{r_s^3r_0}{r^4} \right] H_2^{(1)}(\sqrt{\lambda}r_s) \df r_s
    }{
    \displaystyle \int_{r_0}^\infty \left( \dfrac{r_0}{r_s} \right) H_2^{(1)}(\sqrt{\lambda}r_s) \df r_s
    }
    \right]
    \right)
\end{equation}

\section{Compressible flow solution to the benchmark case}
\label{txt:solutionToTheCompressibleProblem}

The compressible linearised Navier-Stokes equations are considered here to study the flow behaviour in a compressible setting. 
The assumptions made in \citet[{\S} 3]{MorfeySorokinGabard_jfm12} are followed to make the problem analytically tractable, that is, the thermal expansion of the fluid is neglected as well as terms involving gradients of the mean flow. This leads to the set of linearised equation herafter,
\begin{equation}
    \left\{
    \begin{array}{l}
         \dfrac{1}{a_0^2} \dfrac{\partial p}{\partial t} + \rho_0 \nabla \cdot \vc{u} = 0 \\[2ex]
         \dfrac{\partial \vc{u}}{\partial t} + \dfrac{\nabla p}{\rho_0} - \nu \nabla \cdot \left( (\nabla \vc{u}) + (\nabla \vc{u})^T \right) + \left(\dfrac{2}{3}\nu - \dfrac{\mu_b}{\rho_0}\right) \nabla (\nabla \cdot \vc{u}) = \vc{0}
    \end{array}
    \right.
    \label{eq:linearisedNS}
\end{equation}
Considering the time Fourier transform $(p, \vc{u}) = (\hat{p}, \hat{\vc{u}})e^{-i\Omega t}$, the first equation of (\ref{eq:linearisedNS}) is transformed into,
\begin{equation}
\nabla \hat{p}/\rho_0 = a_0^2 \nabla (\nabla \cdot \hat{\vc{u}})/(i \Omega)
\label{eq:ExpressionOfPressure_CompressibleFlow}
\end{equation}
which enables to write an equation solely for $\hat{\vc{u}}$,
\begin{equation}
    -i\Omega \hat{\vc{u}} - \nu \Delta \hat{\vc{u}} - \left( \dfrac{1}{3}\nu + \dfrac{\mu_b}{\rho_0} + \dfrac{a_0^2}{i \Omega} \right) \nabla (\nabla \cdot \hat{\vc{u}}) = \vc{0}
\end{equation}
Unlike in section {\S} \ref{txt:takingAcountBC}, a Helmholtz decomposition in polar coordinates of the velocity field is considered
\begin{equation}
    \hat{\vc{u}} = 
    \left(
    \begin{array}{c}
         \dfrac{\partial \varphi}{\partial r} + \dfrac{1}{r} \dfrac{\partial \psi}{\partial \theta} \\[2ex]
         \dfrac{1}{r} \dfrac{\partial \varphi}{\partial \theta} - \dfrac{\partial \psi}{\partial r}
    \end{array}
    \right)_{(\vc{e}_r,\vc{e}_\theta)}
    \label{eq:VeloPotentialDecompo}
\end{equation}
for which the following relation holds,
\begin{equation}
    \Delta \hat{\vc{u}} =
    \left(
    \begin{array}{c}
         \dfrac{\partial  \Delta\varphi }{ \partial r} 
    + \dfrac{1}{r}
    \dfrac{\partial \Delta\psi}{\partial \theta}
    \\[2ex]
        \dfrac{1}{r} 
    \dfrac{\partial \Delta\varphi}{\partial \theta}
    - \dfrac{\partial  \Delta\psi }{ \partial r} 
    \end{array}
    \right)    
    \hspace{1cm} \text{and} \hspace{1cm}
    \nabla (\nabla \cdot \hat{\vc{u}}) =
    \left(
    \begin{array}{c}
         \dfrac{\partial  \Delta\varphi }{ \partial r} 
    \\[2ex]
        \dfrac{1}{r}
        \dfrac{\partial \Delta\varphi}{\partial \theta}
    \end{array}
    \right)    
    \label{eq:secondOrderSpaceDerivativeWithPotentials}
\end{equation}
where $\Delta$ either refers to the scalar or the vector Laplacian, depending on the context.
Inserting equations (\ref{eq:VeloPotentialDecompo}) in the above system
and after rearranging the equation and introducing $\tilde{\nu} = \dfrac{4}{3}\nu +\dfrac{\mu_b}{\rho_0}+\dfrac{a_0^2}{i\Omega}$, one obtains, without further approximation,
\begin{equation}
    \left\{
    \begin{array}{c}
         \Delta (i\Omega \psi + \nu \Delta\psi) = 0 \\[2ex]
             \Delta \left( i\Omega \varphi + \tilde{\nu} \Delta\varphi \right) = 0
    \end{array}
    \right.
\end{equation}
which correspond to a more concise and slightly more general form as given in equations (D6) and (D7) of \cite{MorfeySorokinGabard_jfm12}.
Following \cite{MorfeySorokinGabard_jfm12} it is assumed in the following that $\Delta\,f=0$ implies $f=0$\footnote{This is debatable especially if the domain of interest $\mathcal{D}$ is bounded. Let us assume $\Delta(f)=0$, then $0=\int_{x\in\mathcal{D}} f \nabla \cdot \nabla f \df x = \int_{x\in\mathcal{D}} \nabla \cdot \left( f\nabla f\right) - ||\nabla f||^2 \df x$. From Ostrogradsky's theorem, $\int_{x\in\mathcal{D}} ||\nabla f||^2 \df x = \int_{x\in\partial\mathcal{D}} f\nabla f \cdot \vc{n}\, \df x$ with $\vc{n}$ the surface normal vector that is pointing outward. For $\nabla f$ to be null and $f$ possibly equal to zero, it appears that $f$ must obey some specific Dirichlet and/or Neumann boundary conditions on $\partial \mathcal{D}$. This is assumed here.
}
It appears that $\varphi$ and $\psi$ are solution of a diffusion equation as for the vorticity field, which is solved step by step in {\S} \ref{eq:GeneralSolVorticityEq}. The general expressions for $\varphi$ and $\psi$ thus are,
\begin{equation}
    \left\{\begin{array}{l}
     \psi(r,\theta) = \displaystyle \sum_{n\in\mathbb{N}} \left( \rho_n e^{in\theta} + \tau_n e^{-in\theta}\right) H_n^{(1)}\left( \sqrt{\dfrac{i\Omega}{\nu}}r\right)
     \\[2ex] 
     \varphi(r,\theta) = \displaystyle \sum_{m\in\mathbb{N}} \left( \xi_m e^{im\theta} + \zeta_m e^{-im\theta}\right) H_m^{(1)}\left( \sqrt{\dfrac{i\Omega}{\tilde{\nu}}}r\right)
    \end{array}\right.
\end{equation}
The boundary conditions for the undeformable oscillating cylinder introduced in {\S} \ref{txt:undeformableCylinder} provide,
\begin{equation}
    \left[ \dfrac{\partial \varphi}{\partial r} + \dfrac{1}{r} \dfrac{\partial \psi}{\partial \theta} \right]_{r=r_0} = 0
    \hspace{1cm} \text{and} \hspace{1cm}
    \left[ \dfrac{1}{r} \dfrac{\partial \varphi}{\partial \theta} - \dfrac{\partial \psi}{\partial r} \right]_{r=r_0} = U
\end{equation}
By making use of $\partial \left(H_n^{(1)}(u)\right)/\partial u = n H_n^{(1)}(u)/u - H_{n+1}^{(1)}(u)$,
and grouping the expressions by azimuthal order, it follows,
\begin{equation}
    (\rho_0 + \tau_0)H_{1}^{(1)}(k_{\nu}r_0) = U
    \hspace{1cm} \text{and} \hspace{1cm}
    \xi_0 + \zeta_0 = 0
\end{equation}
and $\forall n>0$, $\rho_n=0$, $\tau_n=0$, $\xi_n=0$ and $\zeta_n=0$.
Thus,
\begin{equation}
     \psi(r,\theta) =  U \dfrac{H_0^{(1)}\left( \sqrt{\dfrac{i\Omega}{\nu}}r\right)}{H_{1}^{(1)}\left(\sqrt{\dfrac{i\Omega}{\nu}} r_0\right) }
     \hspace{1cm} \text{and} \hspace{1cm}
     \varphi(r,\theta) = 0
\end{equation}
The same expression (\ref{eq:solution_Morfey}) for the incompressible flow problem is finally retrieved with this compressible flow formulation.
However, from equation (\ref{eq:ExpressionOfPressure_CompressibleFlow}), the problem is found to be isobar at this level of approximation.

\bibliographystyle{jfm}
\bibliography{biblio}

\end{document}